\documentclass[aps, twoside, nofootinbib]{revtex4}
\usepackage{amsfonts}
\usepackage{amssymb}
\usepackage[mathscr]{eucal}
\usepackage[dvips]{graphicx}

\def \be{\begin{equation}}
\def \ee{\end{equation}}
\def \bes{\begin{eqnarray}}
\def \ees{\end{eqnarray}}
\def \arr{\rightarrow}

\newcommand{\Cb}{{\rm \bf C}}

\def \sl2{SL(2,\Cb)}

\begin{document}
\title{Causality and matter propagation in 3d spin foam quantum gravity} 
\author{{\bf Daniele Oriti}\footnote{d.oriti@damtp.cam.ac.uk} and {\bf Tamer Tlas}\footnote{t.tlas@damtp.cam.ac.uk}} \affiliation{\small Department of Applied Mathematics and Theoretical Physics, \\ Centre for Mathematical Sciences, University of Cambridge, \\
Wilberforce Road, Cambridge CB3 0WA, UK}
\begin{abstract}
In this paper we tackle the issue of causality in quantum gravity, in the context of 3d spin foam models. We identify the correct procedure for implementing the causality/orientation dependence restriction that reduces the path integral for BF theory to that of quantum gravity in first order form. We construct explicitly the resulting causal spin foam model. We then add matter degrees of freedom to it and construct a causal spin foam model for 3d quantum gravity coupled to matter fields. Finally, we show that the corresponding spin foam amplitudes admit a natural approximation as the Feynman amplitudes of a non-commutative quantum field theory, with the appropriate Feynman propagators weighting the lines of propagation, and that this effective field theory reduces to usual QFT in flat space in the no-gravity limit.
\end{abstract}

\maketitle
\section{Introduction and motivation}
In recent years, spin foam models \cite{review, alexreview,thesis} have attracted much interest as a
new implementation of the sum-over-histories quantization of gravity \cite{teitelboim}
in a discrete setting and in a purely group-theoretic and algebraic
language. Spin foam models can in fact be understood as background independent discrete quantum gravity path integrals: spacetime (and its topology) is represented by a combinatorial 2-complex (collection of vertices, links and faces) and a possible configuration of geometric data is determined by a collection of either (Lorentz) group elements or group representations assigned to its links or faces, respectively; each geometric configuration is assigned a quantum amplitude (usually, but not necessarily, derived via a discretization of some continuum formulation of classical gravity), and the partition function of the model is defined by the typical sum over all possible quantum configurations weighted by the chosen amplitude. The 2-complex is usually chosen to be topologically dual to a simplicial complex of the appropriate dimension.  
The situation as regards 4-dimensional quantum gravity
in a spin foam context is still in rapid evolution, with various models
being proposed and several issues still to be clarified \cite{review, alexreview}. On the other
hand, the viability
of a spin foam quantization of 3d quantum gravity is not anymore an
issue, as it has been shown that the Ponzano-Regge spin foam model
(see for example \cite{PRI}) provides, after suitable regularization, a well-defined and useful description of it, whose
equivalence with other approaches, e.g. Chern-Simons quantization \cite{PRII} or
canonical loop quantum gravity \cite{alexkarim}, can
be established. Moreover, spin foam models can be obtained from the
perturbative expansion of \lq group field theories\rq \cite{ioGFT,laurentGFT,GFTbook}, peculiar
quantum field theories over group manifolds providing a kind of \lq
local simplicial third quantization of gravity\rq, in the same way as
the traditional path integral for quantum gravity can be obtained
(formally) from the perturbative expansion of a suitably defined field
theory on superspace (space of geometries). In doing so, one obtains
also a definite prescription for a sum over topologies, complementing
the sum over geometries provided by the spin foam model (gravity path
integral) alone. Group field theories, as argued in
\cite{ioGFT,laurentGFT,GFTbook}, have a potential and scope that go far
beyond that of spin foam models per se, that make them fascinating, and that
have not been explored but to a very limited extent. \\

The coupling of matter fields to quantum gravity can also be described
in the spin foam setting \cite{danhend,
  cranematter}. In particular, in the 3d case a precise
description of quantum gravity coupled to point particles of any mass
and spin was provided in \cite{PRI,PRII,PRIII}, as a spin foam model, based on the classic work of \cite{hooft,
  matschullwelling,desousa}, and in \cite{iojimmylaurent,iojimmy} in the group field
theory formalism. The basic idea behind the construction of \cite{PRI}
is to treat Feynman graphs of a generic matter field theory as quantum
gravity {\it observables}, and to obtain a description of the interaction of
matter degrees of freedom and quantum gravity ones for each line of propagation from the analysis of the classical coupling of gravity and point particles, with the latter treated as topological defects of the former.     
This led not only to a definition of a spin foam model (and later a group field theory) for 3d quantum gravity coupled to matter, but also to the derivation of an effective non-commutative field theory for scalar fields \cite{PRIII} (and higher spin ones \cite{noispinning}), whose Feynman graphs amplitudes are given by the same quantum amplitudes of the coupled spin foam model, suitably re-written \cite{PRIII}. The amplitudes for the matter Feynman graphs obtained from the coupling with quantum gravity are, however, a bit peculiar, in that they make use of the Hadamard propagator, instead of the Feynman one, to describe the propagation of matter, i.e. they assign the same weights to the external and internal lines of the Feynman graphs, and a rather ad hoc procedure is needed to turn these into the more conventional form \cite{PRIII,PRIIIbis}. This can be interpreted, as we are going to prove, as a result of a missing causality restriction on the {\it quantum gravity} amplitudes one starts with, and that in turn leads to a non-causal propagation of matter degrees of freedom at the quantum level, once these have been coupled to gravity.\\

Motivated partially by this interpretation of the results of \cite{PRIII} and mainly by general considerations on the role of causality in quantum gravity, and in spin foam models in particular (articulated in \cite{causal, feynman, generalised, GFTbook}), we set out to analyze thoroughly the issue of causality in the context of 3d spin foam quantum gravity, with and without matter coupling. While the issue we deal with is indeed very general, and we expect the ideas and techniques we use to deal with it to be general as well, the 3d setting leads to a particularly neat application of them, as we are going to see.\\    

Before presenting the details of our analysis and results, we feel it is appropriate to discuss first the role of causality in quantum gravity in the path integral formalism and spin foam models, and the general arguments for the need for a {\it causality} restriction on the histories summed over\footnote{Causality restrictions, one of which is the one on the lapse/proper time/orientation we refer to here, in a discrete path integral for quantum gravity, together with the consequences they have for the continuum limit of the same, were shown to be crucial also in another approach to discrete quantum gravity, namely Causal Dynamical Triangulations \cite{CDT}.}, and then the strategy for the coupling of matter fields to 3d spin foam models used in \cite{PRI, PRIII}. This will clarify the context, the underlying ideas and the motivations of our work.  

\subsection{Causality in quantum gravity and spin foam models}
The path integral formulation of quantum gravity entails the construction of a partition function of the form:
\[
Z=\int\mathcal{D}g\,e^{iS_M(g)}
\] 
with $S(g)$ being the gravitational Einstein-Hilbert action, for example, for a given spacetime manifold $M$, and the integral being of course over all Lorentzian geometries (metrics up to diffeomorphism equivalence) compatible with the given topology. The resulting quantity is in general a {\it complex} number and changes into its complex conjugate under reversal of the spacetime orientation \cite{teitelboim, hartlehalliwell}. It also allows in principle for the computation of quantum gravity transition amplitudes (using a canonical gravity terminology)

\be
Z(h_1,h_2)=\int_{g \mid h_1,h_2}\mathcal{D}g\,e^{iS_M(g)} \label{eq:path}
\ee
 where $h_1$ and $h_2$ are 3d geometries induced by {\it all} 4d geometries $g$ on the two (disjoint) boundaries of $M$ (corresponding to possible canonical quantum gravity {\it states}). the complex nature and symmetry properties of the above quantities under orientation reversal are the same as those of the partition function. It is obvious that, once defined more rigorously, the partition function and transition amplitudes would encode the full dynamics of the classical gravity theory at the quantum level, for any choice of spacetime topology $M$. Also, the above transition amplitudes would appear, in a third quantized formulation involving spatial topology changing processes \cite{giddingsstrominger}, as the 2-point function (analogue of the Feynman propagator for a single particle) associated to internal lines of any 3rd quantized Feynman graph, confirming that it correctly captures the quantum gravity propagating degrees of freedom (in the cosmological context of \cite{giddingsstrominger} representing the degrees of freedom of the whole \lq quantum universe\rq).\\

A sum-over-histories formulation of the dynamics is possible also in canonical quantum gravity framework. The dynamical content of the theory is encoded in the Hamiltonian constraint operator $\mathcal{H}$ and one relevant object to define for computing physical probability amplitudes, including transition amplitudes, is the inner product between physical states, i.e. solutions of the Wheeler-DeWitt equation $\mathcal{H}\mid \Psi \rangle=0$. One possible way to define such an inner product is in terms of matrix elements of a {\it projector} operator from kinematical states onto physical states (see e.g. \cite{carloprojector} and references therein), to be thought of as some regularized version of $\mathcal{P}=\int \mathcal{D}T e^{iT\mathcal{H}}=\delta(\mathcal{H})$. The sum-over-histories formalism may be used to provide a covariant definition of these matrix elements (including the \lq vacuum-vacuum\rq one, equivalent to the partition function), and a covariant definition of physical wave functions (satisfying all the quantum gravity constraints, including the Hamiltonian one).   
The Lagrangian path integral given above has then an Hamiltonian counterpart in the case of $\Sigma\times \mathbb{R}$ topology, i.e. when a canonical decomposition of the variables is possible. It is given by    

\[
G(h^2_{ij},h^1_{ij})=\int\mathcal{D}N\mathcal{D}N^i\left[
e^{i \int_{\mathcal{M}} d^3x dt \left(\pi^{ij}\dot{h}_{ij} - N
\mathcal{H} - N^i \mathcal{H}_i\right)} \mathcal{D}\pi^{ij}
\mathcal{D}h_{ij}\right].
\]

The (formally defined) integrals are over 3-metrics $h_{ij}$, their conjugate momenta $\pi^{ij}$, over the shift vector $N^i$ (this integration amounts to the imposition of the spatial diffeomorphism constraint) and over the lapse function $N$. The choice of integration range over the lapse is crucial: the choice of the full infinite range $(-\infty,+\infty)$ provides us with the result we were aiming for, i.e. a covariant definition of the physical inner product between canonical quantum gravity states and thus of the projector operator. In fact the resulting quantity $G_H(h^2_{ij},h^1_{ij})\equiv\langle h_2 \mid h_1 \rangle \simeq \langle h_2 \mid \delta(\mathcal{H})\mid h_1\rangle $ is a {\it solution} of the Hamiltonian constraint equation in both its arguments \cite{teitelboim, hartlehalliwell}, and it is a real quantity, as expected from the canonical inner product. However, $G_H(h^2_{ij},h^1_{ij})$ {\it does not} correspond to the lagrangian path integral (\ref{eq:path}); the reason is \cite{hartlehalliwell} that positive and negative lapses correspond to the {\it same} class of 4-geometries, and the difference between the two half-ranges $(0,+\infty)$ and $(-\infty, 0)$ is only that they correspond to {\it opposite spacetime orientations}. In other words, the above quantity in Lagrangian formulation is given by a different {\it symmetric} choice of quantum amplitude, but corresponds to the same set of 4d geometries being integrated over:
\be
G_H(h^2_{ij},h^1_{ij})=\int\mathcal{D}g\left( e^{iS(g)} +
e^{-iS(g)}\right) \label{eq:pathsym}
\ee

A second choice is possible: one may choose to integrate only in the range $(0,+\infty)$ to define:
\[
G_F(h^2_{ij},h^1_{ij})=\int_0^{+\infty}\mathcal{D}N\int\mathcal{D}N^i\left[
e^{i \int_{\mathcal{M}} d^3x dt \left(\pi^{ij}\dot{h}_{ij} - N
\mathcal{H} - N^i \mathcal{H}_i\right)} \mathcal{D}\pi^{ij}
\mathcal{D}h_{ij}\right].
\]

Now, the resulting quantity $G_F(h_1,h_2)$ is indeed {\it equal} to the Lagrangian path integral and so it also captures the full dynamical content of the theory; however, it {\it does not} satisfy the Hamiltonian constraint equation in either of its arguments; it is instead \cite{teitelboim, hartlehalliwell} a Green function for the Hamiltonian constraint operator, satisfying $\mathcal{H}G_F(h_1,h_2)= \delta(h_1 - h_2)$. It defines thus a quantity analogous to $\langle h_2 \mid \frac{1}{\mathcal{H}}\mid h_1\rangle$.\\

This apparently puzzling situation can be easily understood as arising from the difference between Lagrangian and Hamiltonian symmetries, the first being the 4d spacetime diffeomorphisms and the second being the transformations generated by the canonical operators $\mathcal{H}_i$ and $\mathcal{H}$ \cite{hartlehalliwell}. The second set of symmetries is actually {\it larger} than the first and in particular the range $N\in (0,+\infty)$ is perfectly symmetric under transformations of the lapse corresponding to 4d spacetime diffeos, while it is not under canonical symmetries that can connect positive and negative lapses, thus requiring a symmetric range $N\in (-\infty, +\infty)$. Therefore, while the Lagrangian path integral $Z(h_1,h_2)= G_F(h^2_{ij},h^1_{ij})$ is fully symmetric under spacetime diffeos, it is {\it not} symmetric under the transformations generated by the Hamiltonian constraint and one needs a further symmetrization (\ref{eq:pathsym}) to satisfy the Wheeler-DeWitt equation.\\ Another way to see what is going on is to realize that all this has a very precise analogue in the sum-over-histories formulation of the dynamics of a relativistic particle, with $G_H(h^2_{ij},h^1_{ij})$ corresponding to the Hadamard propagator, given in momentum space by $\delta(p^2+m^2)$, thus imposing the Hamiltonian constraint equation $p^2 +m^2=0$, and $G_F(h^2_{ij},h^1_{ij})$ corresponding instead to the Feynman propagator, given in momentum space by $\frac{i}{p^2+m^2 + i\epsilon}$, thus relaxing the same Hamiltonian constraint at the quantum level. 

\medskip

The restriction to positive lapses is a {\bf causality} restriction \cite{teitelboim} in that it corresponds to imposing that the final 3-geometry $h_2$ lies in the causal future of the first $h_1$, and imposes a 'timeless ordering' between initial and final boundary data; again, it has a precise analogue in the relativistic particle case, where it corresponds to a restriction on the particle proper time to positive values only. We want to stress once more another consequence of the restriction, which is the consequent relaxation of the Hamiltonian constraint, or in other words of the classical dynamical condition in gong over to the quantum regime.

The Hamiltonian constraint fully encodes the dynamical content of the classical Einstein's equations for the gravitational field (just as it encodes the Klein-Gordon equations of motion for the relativistic particle). The need to relax this classical constraint is then understood as the need to allow for purely quantum histories, i.e. for virtual propagation. Another way to put it is to say that quantum dynamics takes place {\bf off-shell} with respect to the classical constraint. This is well-known of course in quantum theories of matter dynamics, but it seems not to have been fully realized in the context of quantum gravity dynamics. This is true, however, and accepted for 'interacting systems', as in quantum field theory. The relevance of these considerations for quantum gravity may be less obvious. There are, however, several reasons why we think 'off-shell' dynamics has to be sought for also in quantum gravity, in other words why the Hamiltonian constraint, and the corresponding symmetry, of canonical quantum gravity should {\it not} be imposed necessarily on quantum gravity states and why the dynamics should be allowed to take place 'off-shell' with respect to it. Without entering into much details, let us summarize them.

One reason is the above mentioned wish to impose causality restrictions, or equivalently a notion of causal, timeless ordering among quantum gravity states in computing physical transition amplitudes. Another reason is that, as we have seen, such a restriction from the Hamiltonian perspective is {\bf necessary} to construct a true sum-over-histories formulation of quantum gravity dynamics from the Lagrangian point of view, i.e. a Lagrangian path integral, fully diffeo-invariant, but with no extra symmetry being imposed. Also, the inclusion in the formalism of spatial topology change by going to a 'third quantized' framework \cite{teitelboim, giddingsstrominger} involves indeed the inclusion of interactions among universes and here is where the covariant off-shell path integral appears as a weight for quantum gravity 'internal lines' of propagation, again playing the role the Feynman propagator plays in a usual QFT. One may of course question the physical content of such a quantum cosmological framework, but even these doubts are solved by passing to the spin foam formalism. In fact, on one hand spin foam models are understood as providing a more rigorous definition of the path integral framework for quantum gravity, and thus all the previous considerations regarding causality restrictions and off-shell propagation apply to them as well. On the other hand they also accommodate rather easily the presence of timelike boundaries, thus providing a {\it local} realization of the sum-over-histories idea and thus evade worries about the physical interpretation of the quantum cosmology setting. Finally, the group field theory approach and the role of spin foam models within it realize also the 'third quantization' idea in a discrete and local way, thus calling once more for the need for off-shell propagation of quantum gravity degrees of freedom and for {\it causal spin foam models}, by clarifying in which sense spin foam models deal with 'interacting gravity degrees of freedom', or better put, interacting {\it building blocks of spacetime}, that must thus be allowed to propagate off-shell.

\medskip

Motivated by all of the above, the construction of such causal spin foam models was initiated in \cite{causal}, continued and refined in \cite{feynman} and then in the group field theory formalism in \cite{generalised}. The starting point is the realization that current spin foam models, including the Ponzano-Regge model for 3d quantum gravity, possess the same symmetry under reversal of spacetime orientation that characterizes covariant realizations of the projector operator onto solutions of the Hamiltonian constraint. This explains their providing real amplitudes, as opposed to the complex ones one would expect from a Lagrangian path integral. All this is fine, and indeed spin foam models were first constructed exactly as a way to define in a  covariant language the dynamics of canonical loop quantum gravity, but, as we said, there are several reasons why one needs to go beyond this initial, albeit important, goal. The mentioned symmetry, as appropriate given the discrete setting at the roots of spin foam models, is realized at the level of each dual face of the spin foam 2-complex, as a symmetrization between the two opposite orientations it can be given. This was noticed in \cite{causal} for the 4d Barrett-Crane model together with a consistent way of breaking such symmetry and constructing a possible causal version of the same. This breaking was refined to a more elegant and general procedure based on the particle analogy, that could be applied to any spin foam model in \cite{feynman}, and finally in \cite{generalised} a generalized group field theory formalism was constructed, that allows for the derivation of causal spin foam models, as well as the usual ones, in any dimension and signature, and for a variety of possible quantum spin foam amplitudes. In this process, however, the link with the usual path integral formalism for quantum gravity (continuum and discrete) became less manifest. In this paper, we derive a causal spin foam model for 3d quantum gravity, that represents the causal counterpart of the usual Ponzano-Regge model, starting directly from a discretization of the continuum path integral for gravity in first order formalism, and providing a clear discrete counterpart of the continuum causal restriction on the lapse.

\subsection{Causality in QG coupled to matter: on-shell versus
  off-shell propagation of matter fields}
One additional motivation for imposing a causality restriction on quantum gravity variables is that such a restriction may be in fact {\bf needed} for describing the quantum off-shell causal propagation of matter degrees of freedom of interacting field theories, once these have been coupled to quantum gravity ones and more general spin foam models have been constructed. Recall that one example of this is represented by the recently constructed coupled Ponzano-Regge spin foam model, as we have mentioned and that we are going to review briefly in the following. There are several reason one may expect this at a somehow heuristic level. First of all the dynamics of, say, a relativistic particle coupled in the Hamiltonian context is encoded in the {\it coupled} Hamiltonian constraint for gravity and particle, and it is hard to imagine a procedure (and a motivation for it) that would relax the particle part of it, producing the off-shell propagation coded in the Feynman propagator, without affecting and relaxing constraints on the the pure gravity sector. Second, the restriction on the particle side that gives the Feynman propagator as opposed to the Hadamard one is a restriction on the {\it proper time} of the particle, i.e. the spacetime length of the particle trajectory, which is intrinsically a gravitational variable/observable, and thus the imposition of such a restriction is best interpreted as a restriction on the range of integration of some gravity variables. Even more generally speaking, the very notion of time ordering or causal ordering that is at the root of the definition of the Feynman propagator for a relativistic particle, and of the time-ordered n-point functions of quantum fields, is in fact a {\it quantum gravity} notion, so it must find its origin in the appropriate causality restriction on the quantum gravity side.\\

Once convinced of these rather heuristic motivations, one naturally comes to expect that, conversely, a {\it missing} causality restriction on the quantum gravity side would imply an \lq a-causal\rq and ultimately wrong treatment of particle dynamics, in particular an a-causal propagation of matter degrees of freedom, once these have been coupled to quantum gravity in some spin foam model. From this perspective, then, an apparently puzzling and a rather disturbing feature of the coupled Ponzano-Regge model of \cite{PRI,PRIII} looks instead natural and confirms the above motivations and expectations. The feature we refer to is the fact that, as we are going to see in a few more details in the next section, the Feynman graphs for the effective field theory describing the matter degrees of freedom have Hadamard propagators instead of Feynman ones weighting their internal lines, and thus describe these degrees of freedom as being always on-shell, i.e. always solving the classical Hamiltonian constraint/equations of motion. This also implies a sort a strong a-causality in the sense that no ordering of their arguments is encoded in the Feynman amplitudes (and correspondingly there is no difference between particles and anti-particles). This turns out to be true both at the non-commutative field theory level, encoding the full quantum gravity corrections to matter propagation, and in the no-gravity limit, when one indeed recovers a matter field theory in flat space but, as said, with Hadamard propagators in place of Feynman ones. This means that the result cannot be attributed to any sort of quantum gravity corrections to ordinary matter propagation, but it has really its origin in some more basic feature of the formalism. 

Of course, having the Hadamard propagator on each internal line, one can try to reconstruct from it the corresponding Feynman propagator, and this is indeed done in \cite{PRIII,PRIIIbis}. However, on the one hand this is necessarily a very ad hoc procedure and does not teach us much about the origin of this puzzling feature of the model nor about its meaning, and on the other hand it is not a {\it unique} procedure in that many different functions representing the non-commutative version of the usual Feynman propagator can have the same on-shell (Hadamard) counterpart, because a generic p-dependent factor may become a constant function of the mass m only, due to the $P^2-m^2$ constraint being imposed, as we are going to see and discuss, and because such p-dependent factors may also disappear in the no-gravity limit, and thus such different non-commutative versions of the Feynman propagator may end up having the same flat space counterpart.          

We will see in detail how our {\it ab initio} procedure solves all these ambiguities. The usual coupling of particles to 3d quantum gravity, {\it plus} the {\it same} causality restriction of the pure gravity case, automatically provides us with a well-defined causal version of the coupled Ponzano-Regge model, and with causal Feynman amplitudes for matter fields, with a unique prescription for the non-commutative version of the Feynman propagator, and one that of course has the right and expected no-gravity limit.  

\medskip

\vspace{1cm}

Having presented the motivation for our work, let us now outline our results as we are going to present them.
After a very brief review of the spin foam formulation of 3d quantum gravity coupled to matter fields (section II), we will: 1) identify the correct procedure for implementing a {\it causality} restriction in the spin foam formulation of pure 3d quantum gravity (section III); 2) construct explicitly the resulting {\it causal} spin foam model (section III); 3) add matter degrees of freedom to it and construct a {\it causal} spin foam model for 3d quantum gravity coupled to matter fields\footnote{We deal explicitly with scalar matter only, but none of our procedures or results is significantly affected by the presence of spin degrees of freedom, which simply modifies some of the details. Also, we work in the Riemannian quantum gravity setting, exclusively for simplicity of notation and of mathematical details. We stress however that none of our results or procedures used is affected by a switch to the Lorentzian signature, that causes only a slightly greater complexity of the mathematical details, ultimately due to the non-compactness of the relevant group, i.e. $SL(2,\mathbb{R})$.} (section IV); 4) show that the corresponding spin foam amplitudes admit a natural {\it approximation} as the Feynman amplitudes of a non-commutative field theory, very similar to that obtained in \cite{PRIII}, with the appropriate Feynman propagators being used (section V); 5) show that this effective field theory, and its Feynman amplitudes, reduce to the usual QFT action and Feynman amplitudes in flat space, as they should, in the \lq\lq no gravity limit\rq\rq (section V). We then conclude with some outlook on possible developments of the results we have presented.

\section{Spin foam formulation of 3d quantum gravity coupled to matter fields}
Classical 3-dimensional gravity can be formulated in first order formalism as a BF theory. 

The dynamical variables of the theory are: 
\begin{itemize}
\item an $SU(2)$ connection, denoted by A, which can be locally thought as an $\mathfrak{su}\,$(2)-valued 1-form;
\item an $\mathfrak{su}\,$(2)-valued 1-form, denoted by E\footnote{This is true only locally. Globally it is a section of the bundle associated to the principal SU(2) bundle (over which A is defined), via the adjoint representation.},
\end{itemize}and the action is given by:

\begin{equation}
\label{eq: BF action}
S(E,A,M) = \int_M Tr(E \wedge F(A))
\end{equation}
for a generic spacetime manifold $M$, where $F(A)$ is the curvature of $A$, thus a Lie algebra-valued 2-form, and the trace is taken in the adjoint representation.
The theory is quantized using the path integral method, i.e. by the partition function:

\[
Z(M)\,=\, \int \mathcal{D}E\int\mathcal{D}A\,e^{i\,S(E,A,M)}.
\]

The spin foam representation of the theory is obtained by a simple discretization of the above classical and quantum theory \cite{review, alexreview, thesis, PRI}; of course, the discrete setting helps to a great extent in making mathematical sense of the above functional integrals. 
So, we introduce a triangulation $\Delta$, and its topological dual $\Delta^*$. Keeping in mind that the geometric information of a simplicial complex is fully captured by the set of edge lengths, that the curvature is located on (d-2)-simplices (thus again along edges in 3d), and that $E$ is a 1-form, while $F$ is a 2-form, a natural way to discretize the variables is to do the following:
\begin{itemize}
\item integrate the E field along the edges of the triangulation, obtaining $\mathfrak{su}\,$(2) elements, one for each edge.
\item integrate the F field over the dual faces (i.e. over the faces of the dual cellular complex), obtaining an $\mathfrak{su}\,$(2) element for every dual face (dual to an edge of the triangulation).
\end{itemize}
Given that, under any lattice refining, $\Big ( Holonomy = 1 + Curvature + \ldots \Big )$, a convenient but also appropriate choice is to define the discretized curvature $F$ through the holonomy around the boundary of the same dual face on which it is discretized, which in turn can be obtained from a product of parallel transports (path-ordered exponentials of integrals) of the fundamental connection $A$ along dual edges, that give in the end $SU(2)$ group elements.

Thus in the end \cite{PRI} a natural set of variables for the discretized theory is:
\begin{itemize}
\item an $\mathfrak{su}\,$(2) element for every edge e, denoted by $X_e$.
\item an SU(2) element for every dual edge $e^*$, denoted by $g_{e^*}$. The product of all the $g_{e^*}$'s corresponding to the same e in cyclic order, i.e. the holonomy around a dual face dual to the edge e, will be denoted by $G_e$.
\end{itemize}

With this discretization the action above reads
\begin{equation}
\label{eq:discrete action}
S = \sum_{edges} Tr(X_e \, G_e)
\end{equation}

Now, quantization is straightforward. The partition function is given by\footnote{The factor $\frac{(1+\epsilon(G))}{2}$ is needed in order to get a delta function over SU(2). This point was emphasized also in \cite{PRI}, to which we refer for more details. $\epsilon(G) = sign(Cos(\theta))$ where $\theta$ parametrizes G (see Appendix A).}
\begin{equation}
Z = \prod_e \int dX_e  \, \prod_{e^*} \int dg_{e^*}\, \prod_e \frac{(1+ \epsilon(G_e))}{2} e^{ i \big ( \sum_{e} Tr(X_e \, G_e) \big)} \label{eq:3d}
\end{equation}

The integral over $X_e$ in (\ref{eq:3d}) can be performed\footnote{Note that this is true so easily only in the pure partition function, i.e. not in general if one is computing a generic observable that is a function of the $X_e$ as well as of the $g_{e^*}$.}, to give:
\begin{equation}
Z = \prod_{e^*}\int  dg_{e^*}\, \prod_e \delta\left( G_e\right)  \label{eq:3ddelta}
\end{equation}

Now the delta functions on the group can be expanded in irreducible representations using harmonic analysis on $SU(2)$, and then the integrals over the $g_{e^*}$ can be performed as well, leaving us with the spin foam version of the same partition function \cite{PRI}:

\[
Z = \sum_{\{j_e\}} \prod_{e} (2j_e+1) \prod_{\text{tetrahedra}} \left\{ \begin{array}{ccc} j_{e1} & j_{e2} & j_{e3} \\ j_{e4} & j_{e5} & j_{e6}   \end{array} \right\},
\]
where the quantity $2j_e+1$ is the dimension of the representation $j_e$ labeling the edge $e$ of the triangulation, and the weight for each tetrahedron in the triangulation is given by the $6j$-symbol, a function of the six representations labeling the six edges of the tetrahedron itself. 
This is the so-called Ponzano-Regge spin foam model.\\

The coupling of matter fields is obtained, as anticipated, by treating a full Feynman graph $\Gamma$ of a scalar field theory, with its hidden dependence on geometric variables, as a quantum gravity observable. The coupling between geometric and matter degrees of freedom at each line of propagation of the graph is obtained by a discretization of the continuum action describing the coupling of gravity to relativistic point particles in 3d, with the line of the Feynman graph thus being interpreted as the trajectory of a relativistic particle in a 3d spacetime, and by the subsequent integration over particle data. 

One therefore introduces curves $\gamma_i$'s embedded in the spacetime manifold $M$, where $\gamma_i$ is interpreted to be the worldline of the i-th particle. Collectively, these curves make a 1-dimensional graph $\Gamma$, interpreted as a Feynman graph of a field theory. $\Gamma$ will often be called the 'particle graph' below. The analogue of (\ref{eq: BF action}) is then\footnote{This action was first derived in \cite{desousa}, for a nice discussion of it see the beginning of \cite{branes}.}
\begin{equation}
\label{eq:coupled full}
\tilde{S} = \int_M Tr( E \wedge F) + \sum_i \tilde{m}_i \int_{\gamma_i} Tr[(E + d_A q)\, u J_0 u^{-1}]
\end{equation}
where, q is an $\mathfrak{su}\,$(2)-valued 0-form (function) and u is an SU(2)-valued 0-form. $\tilde{m}_i u J_0 u^{-1}$ is interpreted as the classical (thus already on-shell) momentum of the i-th particle and thus shall be denoted by $p_i$\footnote{$ \tilde{m}_i = 4 \pi G m_i$, where $m_i$ is the mass of the i-th particle and G is the Newton's constant. We shall use in this section units in which $4 \pi G = 1$, and hence will call $\tilde{m}_i$ the mass as well, while it is in fact the deficit angle associated to the mass $m_i$. We shall restore Newton's constant to the equations in section V since we will be interested in the limit $G \arr 0$.}. The equations of motion, which are obtained after varying the E and the q are
\begin{eqnarray}
\label{eqnarray:curvature}
F(A(x)) & = & \sum_i p_i \delta\left( x - x_{\gamma_i}\right) \\
\label{eqnarray: momentum}
d_A p_i &=& 0
\end{eqnarray}

The first equation means that the curvature is flat everywhere except at the locations of the particles where it has a conical singularity. The second equation means that the momentum of a particle is covariantly constant along its worldline.\\
The full continuum partition function is now
\[
Z = \int \mathcal{D}A \, \mathcal{D} E \, \prod_i \big ( \mathcal{D}q_i \, \mathcal{D}p_i \big ) \, e^{i  \tilde{S}}
\]

Note that in the path integral we have something of the form 
\[
Z = \int \mathcal{D}q_i \ldots  e^{i \int_{\gamma_i} q_i d_A p_i + \ldots}
\]

The effect of the integral over $q_i$'s is then simply to impose equations (\ref{eqnarray: momentum}). However, as will be apparent in a moment, the discretization itself, if chosen appropriately, will impose these constraints.\footnote{Essentially, this is because each edge of the triangulation (adapted to the curves, or more precisely to the Feynman graph) will carry a \textit{single} group element, to be considered the discretized analogue of p.} Thus we shall drop the last term in (\ref{eq:coupled full}) and discretize a slightly different but equivalent theory given by:

\begin{equation}
\label{eq:classical coupled}
S = \int_M Tr( E \wedge F) - \sum_i \int_{\gamma_i} Tr(E\, p_i)
\end{equation}

To discretize this action we, as before, introduce a triangulation $\Delta$ and its dual $\Delta^*$. We insist, however, on a triangulation that is adapted to $\Gamma$, i.e. the $\gamma_i$'s coincide with some of the edges of $\Delta$. The discretizations of the E and A fields are given as before. The discrete analogue of $p_i$ is given by $u \, e^{(\tilde{m}_i J_0)} \, u^{-1}$, the group element corresponding to the $\mathfrak{su}\,$(2) element $(\tilde{m}_i \, u \, J_0 \, u^{-1})$ \cite{PRI}. It is important to note that there is one u for each edge of $\Gamma$, i.e. the discretized $p_i$ is constant along each edge of $\Gamma$, thus satisfying (\ref{eqnarray: momentum}). Since there is exactly one $\tilde{m}_i$ associated to every edge of $\Gamma$, from here on we shall label the mass of the particle by the edge that it is propagating on.\\

With these definitions, the analogue of (\ref{eq:discrete action}) for action (\ref{eq:classical coupled}) is given by\footnote{The addition of $\mathfrak{su}\,$(2) elements F and $p_i$ is converted into product of SU(2) elements $G_e$ and $u_e^{-1} e^{(\tilde{m}_e J_0)} u_e^{}$.}

\begin{equation}
\label{eq:coupled discretized}
S = \sum_{edges \, \notin \Gamma} Tr(X_e \, G_e) +  \sum_{edges \, \in \Gamma} Tr(X_e \, G_e \, u_e^{-1} e^{(\tilde{m}_e J_0)} u_e^{})
\end{equation}

It is obvious that if one calculates the equations of motion associated to the $X_e$ variables on the particle graph one obtains $G_e = u_e^{} e^{(\tilde{m}_e J_0)} u_e^{-1}\equiv u_e h_e u_e^{-1} $, which is essentially the holonomy version of (\ref{eqnarray:curvature}).\\

The partition function for the coupled theory is analogous to (\ref{eq:3d}):

\be
Z = \prod_e \int dX_e  \, \prod_{e^*} \int dg_{e^*}\, \prod_e \frac{(1+\epsilon(G_e))}{2} \,  \prod_{e\in\Gamma}\int d u_e \,e^{ i \big ( \sum_{e\notin\Gamma} Tr(X_e \, G_e)\,+\, \sum_{e\in\Gamma} Tr(X_e \, G_e\,u_e^{} e^{(\tilde{m}_e J_0)} u_e^{-1}) \big)} \label{eq:3dparticle}
\ee

Also in this case the integral over $X_e$ can be performed explicitly (again, for observables that are not functions of them) to give:
\be
Z = \prod_{e^*}\int  dg_{e^*}\,\prod_{e\in \Gamma}\int du_e \prod_{e\notin \Gamma} \delta\left( G_e\right)\prod_{e\in\Gamma} \delta\left( G_e u_e h_e u_e^{-1}\right)  \label{eq:3ddeltaparticle},
\ee
and also in this case one can obtain a spin foam presentation of the same partition function using the harmonic analysis on $SU(2)$ and thus expanding the delta functions in (\ref{eq:3ddeltaparticle}) in irreps:

\[
Z = \sum_{\{j_e\}} \prod_{e\notin\Gamma} (2j_e+1) \prod_{e\in\Gamma}\chi^{j_e}(h_e) \prod_{\text{tetrahedra}} \left\{ \begin{array}{ccc} j_{e1} & j_{e2} & j_{e3} \\ j_{e4} & j_{e5} & j_{e6}   \end{array} \right\}.
\]

This is the so-called coupled Ponzano-Regge model \cite{PRI}, in the case of massive spinless point particles, modulo some arbitrary (at this stage) normalization factor given by a function of the mass variables $\tilde{m}_e$ \cite{PRI,PRIII}.

One key result, obtained in \cite{PRIII}, is that, having re-interpreted the $G_e$ at the location of the particles as their gravity-induced physical momenta at the quantum level, the above partition function can be re-expressed as the Feynman evaluation of a Feynman graph for a non-commutative field theory. The way this is achieved is by introducing non-commutative momenta $P_e$ valued in $\mathcal{B}_\kappa(\mathbb{R}^3)$, i.e. the 3-ball in $\mathbb{R}^3$ with radius $1/\kappa = 4\pi G$, through: $g = \sqrt{1 - \kappa^2 | \vec{P} |^2}1 + \kappa \vec{P} \cdot \vec{J}$, the star product $
e^{\frac{i}{2 \kappa} Tr(X g_1)} \star e^{\frac{i}{2 \kappa} Tr (X g_2)} = e^{\frac{i}{2 \kappa} Tr(X g_1 g_2)} $
and the 'Fourier' transform naturally associated to the above product $f(X) = \int d g \, e^{- i Tr(X P(g))} \tilde{f}(P(g))$. The result (modulo pre-factors) is:
\[
Z \propto \prod_{e\in\Gamma} \int \frac{\kappa^3 d^3 \vec{P}_e}{\pi^2} \prod_{e\in\Gamma}\delta\left( |\vec{P}_e|^2 - \frac{sin^2(m_e \kappa)}{\kappa^2}\right)\, \prod_{v\in\Gamma}\delta{\big(\oplus_{e \in \partial v} \vec{P}_e \big)}.
\]
Basically, the delta function enforcing the conservation of the non-commutative momenta at the vertices of the Feynman graph results from the delta functions for pure gravity edges of the triangulation in (\ref{eq:3ddeltaparticle}), while the {\it Hadamard propagators} for the lines of the particle Feynman graph result from the delta function coupling gravity and matter degrees of freedom in (\ref{eq:3ddeltaparticle}). The problem we had anticipated is now manifest. Hadamard propagators are {\it not} the correct way to capture the quantum dynamics of matter degrees of freedom, in particular they possess the wrong causal properties, and are {\it not} what comes out of the perturbative expansion of a QFT for matter fields. The problem is, as we said, bypassed in \cite{PRIII,PRIIIbis} by re-constructing a possible definition of the Feynman propagator from the obtained Hadamard propagator, and using this possible definition to reconstruct the non-commutative QFT action that would originate the above Feynman amplitudes. However, as mentioned in the introduction, the reconstruction is highly non-unique, in particular because the definition of the Hadamard propagator is non-unique: the insertion of any function of $P_e^2$ is turned by the delta function into an irrelevant insertion of a function of $m_e$. Also, the choice made in \cite{PRIII,PRIIIbis} is natural and gives the correct result in the no-gravity limit, i.e. $G\rightarrow 0$, but, again, it is clear that this limit does not constrain the initial choice enough to achieve uniqueness: any insertion of a function of $P_e^2$ that gives a constant in the limit does not affect the result.     
We will see how a correct {\it ab initio} derivation of a {\it causal} coupled model will solve naturally and uniquely all these difficulties.   

\section{Pure gravity - causal Ponzano-Regge model}
We want now to derive a causal version of the Ponzano-Regge model for pure 3d quantum gravity, starting from the discrete formulation of 3d BF theory reviewed above, and imposing on it the causality restriction that reduces the corresponding path integral to that of 3d quantum gravity, in Lagrangian form. By deriving everything directly from the discretization of the continuum theory, we will avoid any sort of ambiguity (apart those intrinsic to any discretization procedure) and obtain a {\it unique} prescription for how the causal Ponzano-Regge model should differ from the usual one.
   
Let us now go back to the partition function:
\[
Z(M)\,=\, \int \mathcal{D}E\int\mathcal{D}A\,e^{i\,S(E,A,M)},
\]
and its discrete analogue (\ref{eq:3d}):

\begin{equation}
Z = \prod_e \int dX_e  \, \prod_{e^*} \int dg_{e^*}\, \prod_e \frac{(1+ \epsilon(G_e))}{2} \, e^{ i \big ( \sum_{e} Tr(X_e \, G_e) \big)}
\end{equation}

Another fruitful way of looking at it is to consider the action in (\ref{eq:3d}) a discrete analogue of
\begin{equation}
\label{eq:gravity hamiltonian}
\int dt \int d^2 x (N^i H_i + N H)
\end{equation}

Where $N^i , N$ are the shift and the lapse; and $H_i^e , H^e$ are the diffeo and the Hamiltonian constraints. This is almost immediate from (\ref{eq:3ddelta}), since after one integrates over the $X_e$'s one is left with $\delta(G_e)$'s which are essentially imposing the constraints $F_e = 0$, or $H_i^e = 0 = H^e$. This line of reasoning has been explored in detail in \cite{alexkarim}, and indeed it was verified that the above partition function imposes on the spin network states the diffeomorphism as well as the Hamiltonian constraints.\\

However, for reasons mentioned above, we do \textit{not} want to impose the Hamiltonian constraint. Or, put differently, we want to obtain the \textit{causal} analogue of the above partition function.
At the Lagrangian level this means restricting the geometric variables in the partition function to those corresponding to a positive orientation of spacetime only, i.e. constructing a discrete analogue of the constrained partition function:
\[
Z(M)\,=\, \int \mathcal{D}E\int\mathcal{D}A\,\,\Theta\left( det(E)\right)\,e^{i\,S(E,A,M)},
\]
where the $\Theta(det(E))$ is the Heaviside step function effectively imposing the constraint $det(E) \geq 0$, i.e. the positive orientation of spacetime as encoded in the triad. 
Before we perform such a causal modification of the Ponzano-Regge model, let us briefly review its symmetries \cite{diffeos}. In the continuum, PR has two symmetries:
\begin{itemize}
\item The usual SU(2) gauge transformations, given locally by $E \arr g E g^{-1}$ and $A \arr g A g^{-1} + g d g^{-1}$, where g is an SU(2) group element. This symmetry will be called  'rotational' symmetry below.
\item An additional symmetry, which can be seen as originating at the classical level from the Bianchi identity, $E \arr E + d_{A} \phi$, where $\phi$ is an $\mathfrak{su}\,$(2)-valued function. This symmetry will be called a 'translational' symmetry below. \end{itemize}
The above two symmetries combined, are equivalent to diffeomorphism symmetry \underline{\textit{on shell}}.\\

Let us write $X_e = \vec{x}_e \cdot \vec{J}$ and $G_e = Cos(\theta_e) 1 + \vec{n}_e \cdot \vec{J} Sin(\theta_e)$, where the $J$'s are the $\mathfrak{su}\,$(2) generators and $\theta_e$ and $\vec{n}_e$ are the standard parameters for the SU(2) group.\footnote{$\vec{x}_e$ is an element of $\mathbb{R}^3$, $\vec{n}_e$ is an element of $S^2$, and $\theta_e$ is a real parameter which goes between 0 and $\pi$. Check Appendix A for more details.} With these definitions, action (\ref{eq:discrete action}) becomes
\begin{equation}
\label{eq:discrete action vector}
S = - \sum_{edges} \vec{x}_e \cdot \vec{n}_e \, Sin(\theta_e)
\end{equation}
Now, the symmetries become
\begin{itemize}
\item Simultaneous rotations of $\vec{x}_e$'s and $\vec{n}_e$'s, i.e. $\vec{x}_e \arr O_e \vec{x}_e$ and $\vec{n}_e \arr O_e \vec{n}_e$, where $O_e$ is an orthogonal transformation associated to every edge. This is the reason for calling this symmetry rotational.
\item The translational symmetry has a more complicated expression \cite{diffeos}, but essentially it is $\vec{x}_e \arr \vec{x}_e + \vec{\phi}_e$.
\end{itemize}

It should be clear from the discussion in the introduction that the way to impose causality is to put some restriction on the $\vec{x}_e$ integrations. This is because we would like to restrict the integration to 'positive' triads, and triads correspond to $\vec{x}_e$'s after discretization. Obviously, a restriction to the positive triads will not affect the rotational invariance, since it is impossible to convert a positive triad to a negative one by a rotational gauge transformation. Thus, the causal theory has to still be invariant under the rotational symmetry. On the other hand, the translational symmetry has to be broken, at least partially, since it is clear that one \textit{can} go from 'positive' triads to negative ones by doing a translational transformation (just choose $d_A \phi = - 2 E$).\\

Another way of looking at it is to say that we want to restrict to positive lapses only. Now, comparing (\ref{eq:gravity hamiltonian}) to (\ref{eq:discrete action vector}) one could identify $(N,N_i)$ with $\vec{x}$ and $(H,H_i)$ with $\vec{n} Sin(\theta)$. Thus, the translational symmetry implies symmetry under $N \arr N+constant$. So, it is evident that there's no way to reconcile this with the fact that we want $N \geq 0$.\\

So, how should we restrict the range of $\vec{x}_e$'s, to implement the restriction $det E \geq 0$, while keeping the rotational invariance? Before the prescription is provided, note the following four points:
\begin{itemize}
\item[\textbf{a} -] In the continuum, if $E \arr -E$ then $det (E) \arr - det (E)$. Hence, if $\vec{x}_e$ can be thought of as the discretization of a positive triad, then $-\vec{x}_e$ should be considered the discrete analogue of a negative one. Thus, if a certain subset of $\mathbb{R}^3$ corresponds to positive triads, its parity-inverted image through the origin corresponds to the negative ones.
\item[\textbf{b} -] In the continuum, if $det (E) \geq 0$ then $det (\lambda E) \geq 0$ as well, for any $\lambda \geq 0$. Hence, if $\vec{x}_e$ corresponds to positive triads, $\vec{\lambda x}$ does so as well.
\item[\textbf{c} -] The rotational invariance implies that if $\vec{x}_e$ is a positive triad, then $O_e \, \vec{x}$ is also a positive one, where $O_e$ is any rotation with $\vec{n}_e$ as its axis. Note that [\textbf{b}] and [\textbf{c}] taken together imply that, the region of all $\vec{x}$'s which correspond to positive triads is a union of cylindrical cones with axis $\vec{n}$.
\item[\textbf{d} -] Finally consider the continuum action, with a positively oriented triad. Then start 'deforming' the triad until it becomes negative. Since, the Lagrangian is proportional to the volume form, this means that the Lagrangian, switches sign as $E$ is deformed into $-E$. This implies that the Lagrangian passes through zero, along the 'path of deformation'. This implies that if we start from the discrete Lagrangian and start 'deforming' the $\vec{x}$, then as long as the lagrangian doesn't vanish, we haven't switched from positive triads to negative ones. The earliest we can do so is the moment the Lagrangian passes through zero.
\end{itemize}

Thus, assume that $\vec{x}$ is almost along the corresponding $\vec{n}$, and that this $\vec{x}$ corresponds to a positive triad. Then, [\textbf{b}] and [\textbf{c}] imply that all $\vec{x}$'s within a very 'thin' cone with axis $\vec{n}$ are positive as well. Keeping in mind [\textbf{d}], start 'opening' the cone. Then, as is evident from (\ref{eq:discrete action vector}), the earliest the Lagrangian can become zero is when $\vec{x}$ is orthogonal to $\vec{n}$. However, recalling [\textbf{a}], one sees that this is the largest allowed region of positive triads (otherwise the regions where the non degenerate triads are positive and where they are negative overlap).\\
The restriction to positively oriented triads is obtained therefore by considering only those $\vec{x}$ that make a positive dot product with $\vec{n}$. In other words, one has to impose that the projection of the triad along the {\it normal} to the surface to which the holonomy defining the curvature refers is positive. Notice that this is exactly analogous (although a proper Hamiltonian treatment of the discrete action we are dealing with would take us too far) to the continuum restriction on the lapse function, defined indeed as the projection of the triad field along the normal to the surface to which canonical variables refer.

We arrive at the following prescription:
\\
\\
\textbf{\textit{In the partition function (\ref{eq:3d}) one has to integrate only over those $\vec{x}_e$'s such that $\vec{x}_e \, \cdot \, \vec{n}_e \geq 0$}}
\\
\\

If the above geometric arguments were not sufficient to derive and motivate our prescription for the causality/orientation dependence restriction, this could have been motivated also by comparing the action (\ref{eq:discrete action vector}) to the Regge action in first order form, in terms of group-theoretic variables, derived in \cite{magnea}. The expression is:
\begin{equation}
\label{eq:magnea}
S = \sum_{edges} L_e \, Sin(\Theta_e)
\end{equation}
Where, $L_e$ is the length of the edge e, and $\Theta_e$ is the deficit angle at this edge. Comparing (\ref{eq:magnea}) to (\ref{eq:discrete action vector}), and identifying $L_e$ with $\vec{x}_e \, \cdot \vec{n}_e$, it is immediate that the above prescription corresponds to integrating only over positively oriented edge lengths, exactly as one would expect\footnote{\cite{magnea} considered the SO(3) theory, while we are considering the SU(2) one. This is the reason that $\theta_e$ in (\ref{eq:discrete action vector}) is \textit{not} $\Theta_e$ in (\ref{eq:magnea}). In fact $\Theta_e = 2 \, \theta_e$. Classically the two theories are, of course, the same, since the equations of motion, in both cases, imply that $\Theta_e=0=\theta_e$. Note that $\theta_e \in [0,\pi]$, consequently the deficit angle $\in [0, 2 \pi]$. This should be clear, since after parallel transporting a vector around the edge, there's no difference between a rotation by angle $\theta$ or by $\theta + 2 \pi n$. The fact that in the usual Regge calculus the range of the deficit angle is $[-\infty, 2\pi]$ is an outcome of the traditional second order formulation.}.\\

We shall perform the $\vec{x}_e$ integrals in (\ref{eq:3d}). Since the partition function factorizes per edge (dual face), and gives the same weight to each, it is sufficient to do only one of these integrals, for an arbitrary choice of edge, henceforth we shall drop the subscript $e$ from our vectors.\\

So, let us now compute the integral over $\vec{x}$:

\begin{eqnarray}
\label{eqnarray:distro}
\int_{x \cdot n \geq 0} d^3 \vec{x} \, \, e^{- i ( \vec{x} \cdot \vec{n} Sin(\theta))} &=&  \int_{x \cdot n \geq 0} d^3 \vec{x} \, \,  e^{- i \vec{x} \cdot \vec{n} (Sin(\theta) + i \epsilon)} \\
&=&  \int e^{- i r cos(\alpha) (Sin(\theta) + i \epsilon)} \,\, r^2 \, sin(\alpha) dr d\alpha d\beta  \nonumber \\
&=&  \frac{- 2 \pi i }{Sin(\theta) + i \epsilon} \int [e^{i r \gamma (Sin(\theta) + i \epsilon)} - e^{i r (Sin(\theta)+ i \epsilon)}] \, r \, dr \nonumber \\
&=&   \frac{- 2 \pi i}{Sin(\theta)+ i \epsilon} \bigg [ \Big (\frac{1}{i \gamma (Sin(\theta) + i \epsilon)}\Big )^2 - \Big (\frac{1}{i(Sin(\theta)+i \epsilon)} \Big)^2 \bigg ] \nonumber \\
&& \nonumber \\
&=&   \frac{- 2 \pi i}{(Sin(\theta) + i \epsilon)^3}\Big (1-\frac{1}{\gamma^2} \Big ) \nonumber \\
\label{eqnarray:distro2}
& = &  \frac{C_{\delta}}{4}   \frac{i}{(Sin(\theta) + i \epsilon)^3}
\end{eqnarray}
\\

In the above $\gamma$ is the cosine of the upper limit of the $\alpha$ integration, $\gamma = Cos(\frac{\pi}{2} - \delta)=Sin(\delta)$, where $\delta$ is a small positive number, the factor of a quarter is included to recover the usual normalization used in the Ponzano-Regge model if we take the sum of (\ref{eqnarray:distro2}) and its complex conjugate. Due to the importance of the above result, we give a more rigorous derivation of it in Appendix B. The regulator $\epsilon$ plays exactly the same role as the regulator in the Feynman propagator in the usual quantum field theory, as will be apparent in section V. It is understood that it has to be sent to zero at the end of every calculation, i.e. when computing {\it normalized} expectation values of observables, including field correlations. Note that the presence of a small explicit regulator $\epsilon$ is fully equivalent to having $\theta$ analytically continued to the complex domain and having acquired a small positive imaginary part. \\

In the above $\delta$ has to be set to $0$, but it is clear that this gives an infinity. Fortunately, all the infinity is contained in a constant which factors out and does not contribute to the physical calculations we will be interested in. Let us clarify: we will get this constant (which we have denoted by $C_{\delta}$) for every edge of our triangulation, thus the partition function (which becomes a function of $\delta$) will be given by

\[
Z_{\delta} = (C_{\delta})^{|E|} \, Z
\]

where the above equation defines Z, and $|E|$ is the number of edges in the triangulation.\\
Now as far as observables of the connection (A in the continuum, g's in the discrete) are concerned we can as well use Z for calculations, which is independent of $\delta$. This is because as far as the connection is concerned the difference between the different $\delta$'s is a connection-independent constant and thus will cancel from any calculation\footnote{As is well known, for any observable O(A), the expectation value is given by $\langle O(A) \rangle = \frac{\int \mathcal{D}A \mathcal{D}E O(A) e^{iS}}{\int \mathcal{D}A \mathcal{D}E e^{iS}}$. Thus the overall connection-independent constant will be irrelevant for such calculations.}. Since in this paper we will be interested only in connection-related questions, we will use Z from now on.\\
 
What about the observables which are functions of the $\vec{x}$'s? In that case we of course cannot use the above argument and will have to integrate all the way to $\delta = 0$ for those $x$ in the argument of the observable. Whether the expectation value of such observable converges or not is a question which depends on the observable under consideration, and cannot be answered a priori\footnote{The observable should go to zero faster than $\frac{1}{|\vec{x}|^2}$ as $|\vec{x}| \arr \infty$.}.
\\

Keeping equation (\ref{eqnarray:distro2}) in mind, the partition function (\ref{eq:3d}) becomes
\begin{equation}
\label{eq:pure partition}
Z = \int \prod_{e^* \in \Delta^*}   d g_{e^*} \prod_e \frac{(1+ \epsilon(G_e))}{2}  \prod_{e \in \Delta} \frac{1}{4} \frac{i}{(Sin(\theta_e) + i \epsilon_e)^3}
\end{equation}
\\

One could expand now the above expression in characters, then proceed in the usual way, by integrating over $g_{e^*}$'s and obtaining 6-j symbols. The only difference from the usual Ponzano-Regge partition function now would be in the weight assigned to each edge (dual face). So, if the dual face is labeled by a representation j, instead of the dimension of the representation one gets more complicated functions of j's and $\epsilon$'s, denoted by $c_j(\epsilon)$. Consequently, the spinfoam version of (\ref{eq:pure partition}) is given by
\[
Z = \sum_{labellings} \, \, \prod_{e \in \Delta} c_{j}(\epsilon) \prod_{tetrahedra \in \Delta} \{ 6j \}
\]

The sum goes over all labellings of the faces of the dual two-complex by representations of SU(2). In the above $\{ 6j \}$ denotes the usual 6-j symbol, and the $c_{j}(\epsilon)$'s are given by the following formulae  
\begin{eqnarray*}
\lefteqn{\hspace{2.5cm}\frac{(1 + \epsilon(G))}{2} \frac{1}{4} \frac{i}{(Sin(\theta) + i \epsilon)^3}  =  \sum_j \, c_j
  \chi_j(\theta)} \\ 
&& \chi_j(\theta)  =  \frac{Sin( (2j + 1) \theta)}{Sin(\theta)} \;\;\;\;\;\;
c_j(\epsilon)  =  \frac{2}{\pi}  \int_0^{\frac{\pi}{2}} \frac{1}{4} \frac{i}{(Sin(\theta) +
  i \epsilon)^3} \, \chi_j(\theta)  \, Sin^2(\theta) \, d \theta 
\end{eqnarray*}

The coefficients $c_j(\epsilon)$ can be given as linear combinations
of hypergeometric functions.

Going back to (\ref{eq:pure partition}) we note the following properties of this expression:
\begin{itemize}
\item[I] The amplitude is sharply peaked around the classical, on-shell, configuration, $\theta_e=0$. Exactly as one would expect. Heuristically, one can think of the edge amplitudes as 'smoothed' or 'regularized' Dirac delta functions, just as the Feynman propagator for a spinless particle $\frac{i}{p^2-m^2}$ can be considered as a smoothed $\delta(p^2-m^2)$.\footnote{The Feynman propagator is equal to a delta function + a principal value distribution, which can be thought of as a quantum correction to the 'classical' delta.}
\item[II] The amplitude generically is a complex number, moreover $Z \arr \bar{Z}$ when the orientations of the $\vec{x}_e$'s are reversed. This is is not obvious from (\ref{eq:pure partition}) since the $\vec{x}_e$'s have been integrated out. However it is clear from the calculation on the previous page, that if one replaces $\vec{x}$ with $-\vec{x}$ one gets the complex conjugate of the final answer. Again, this is exactly what one would expect from the considerations about the continuum path integral, and matches the result we were aiming to obtain, as discussed in the introduction.
\item[III] The amplitude is \textit{not} triangulation independent. Although this may seem surprising, in fact it should be expected, since we are not on shell anymore.\\One expects triangulation independence because of the fact that classically there are no local degrees of freedom, however to show that this is indeed the case one needs to invoke the equations of motion (constraints) in order to show that there are no local degrees of freedom left after gauging away the symmetries. Since we are not imposing any constraints we have local 'virtual' degrees of freedom, even though classically the theory is totally trivial. This is just like the case of a relativistic particle. Classically the magnitude of the momentum is fixed to be the mass, but quantum mechanically it can be anything.\\
Another way of looking at this, and to understand how no modification (and no restriction) of the BF partition function, and in particular no realization of the Lagrangian path integral for pure gravity, could be triangulation independent, comes from the canonical picture, along the lines introduced in \cite{alexkarim}. If one tries to write an 'evolution' operator U, as opposed to a projection operator onto solutions of the Hamiltonian constraint, using the same approach\footnote{Essentially, one must define some regularized version of $\langle \vert U \vert \rangle = \langle \vert \int d \tau e^{i \tau H} \vert \rangle = \langle \vert  \frac{1}{H+i \epsilon} \vert \rangle$, where H is the Hamiltonian constraint.}, then the rotational gauge invariance implies that U has to be expandable in characters. However the 'fusion' identity will only be satisfied, as proven in \cite{alexkarim}, if $c_j= 2j + 1$, i.e. for the delta function over the group, imposing the Hamiltonian constraint, as in the usual Ponzano-Regge model, or $c_j=0$, thus giving a trivial operator. Without the fusion identity, there can be no regularization independence and hence no triangulation independence.
\item[IV] In the above, we have totally glossed over the issue of gauge fixing \cite{diffeos}. We have mentioned above that the translational symmetry is broken by our restriction on the $\vec{x}$ variables. However, this is only partially true. Roughly, the translations along the $\vec{n}$ are broken, while the ones in the directions orthogonal to $\vec{n}$ remain intact. Thus, what one should do, is to choose a maximal tree T, choose a gauge fixing condition to be implemented along this tree and compute that Faddeev-Popov (FP) determinant. If we choose the simplest gauge condition, which is $(\vec{x} - \vec{x} \cdot \vec{n} = 0)$, i.e. we set the the component of $\vec{x}$ orthogonal to $\vec{n}$ to zero, then using the results of \cite{diffeos} it is easy to show that the FP determinant has the form $\Delta = \prod_{e \in T} ( 1 + | \Omega_e |^2)$, where $\Omega_e$ is a complicated function of all the $G_e$'s associated to the edge e\footnote{Effectively it is given by Baker-Campbell-Hausdorff formula. What we mean by 'the $G_e$'s associated' is the set of all $G_e$'s which correspond to all the edges meeting at the vertex of the original edge. See \cite{diffeos} for details. The only difference from the formulas given in \cite{diffeos} is that our $\Omega_e$ is the projection of the one defined there on the plane orthogonal to $\vec{n}_e$ associated to the edge e.}. In this choice of gauge fixing, the amplitudes for gauge fixed edges are given by a regularized $\frac{i}{\sin{\theta}+i\epsilon}$ instead of the cubic inverse power of the un-gauge-fixed amplitudes, as it is easy to verify by direct calculation. \\
It can be shown that in the case when we have delta functions everywhere as edge amplitudes, i.e. by using the {\it classical} equation of motion, then the FP determinant evaluates to one. In our case this is no longer the case, since our edge amplitudes have support away from the flat connections. Since the FP determinant is a nonlocal function (what we mean by this is that it cannot be factorized into a product of contributions each of which is a function of variables coming from a single edge), it depends on the triangulation and does not simplify. The calculation of this Faddeev-Popov determinant, and thus the presentation of the fully gauge fixed causal partition function, will be left for future work.

\end{itemize}

\section{Matter coupling - causal coupled Ponzano-Regge model}
We now consider the case of matter coupling, and show that doing a restriction similar to the one done above one obtains a model similar to the one proposed in \cite{PRIII}, but with the correct Feynman propagators instead of the Hadamard ones obtained there.

We shall now couple gravity to matter along the lines introduced in \cite{PRI}, and reviewed in section II. So we end up with the partition function:

\[
Z = \prod_e \int dX_e  \, \prod_{e^*} \int dg_{e^*}\, \prod_e \frac{(1+ \epsilon(G_e))}{2}  \prod_{e\in\Gamma}\int d u_e \,e^{ i \big ( \sum_{e\notin\Gamma} Tr(X_e \, G_e)\,+\, \sum_{e\in\Gamma} Tr(X_e \, G_e\,u_e^{} e^{(\tilde{m}_e J_0)} u_e^{-1}) \big)}
\]

Once more, before showing what type of causality constraint we are going to impose on this theory and the result of this imposition, let us discuss very briefly the symmetries of the above theory, because it will help in clarifying some aspects of the causal restriction and of the resulting causal model.

The theory retains the symmetries of the pure gravity case, i.e. the rotational and translational ones, if one augments the previous transformations with:
\begin{itemize}
\item $q \arr g q g^{-1}$, and $u \arr  u g^{-1}$, or equivalently $p \arr g p g^{-1}$, for the rotational symmetry.
\item $q \arr q - \phi$, for the translational one.
\end{itemize}

Before we give the partition function, let us write our variables, analogously to the previous section, as:
\begin{eqnarray}
\label{eqnarray:x field}
X_e &=&  \vec{x}_e \cdot \vec{J}\\
\label{eqnarray:group element}
G_e &=& Cos(\theta)1+ \vec{n}_e \cdot \vec{J} Sin(\theta)\\
\label{eqnarray:mass element}
u_e^{-1} e^{(\tilde{m}_i J_0)} u_e^{} &=& Cos(\tilde{m}_e)1+ \vec{u}_e \cdot \vec{J} Sin(\tilde{m}_e)
\end{eqnarray}

With these definitions, the action (\ref{eq:coupled discretized}) becomes
\begin{equation}
\label{eq:traced}
S = \underbrace{ - \sum_{e \notin \Gamma} \vec{x}_e \cdot \vec{n}_e Sin(\theta_e) }_{S_{pure \, gravity} = S_p}  \underbrace{- \sum_{e \in \Gamma}\Big [ Cos(\tilde{m}_e) Sin(\theta_e) \vec{x}_e \cdot \vec{n}_e + Sin(\tilde{m}_e) Cos(\theta_e) \vec{x}_e \cdot \vec{u}_e - Sin(\tilde{m}_e) Sin(\theta_e) \vec{x}_e \cdot ( \vec{n}_e \times \vec{u}_e) \Big ] }_{S_{gravity \, coupled \, to \, matter} = S_c}
\end{equation}

The partition function is
\begin{eqnarray*}
Z & = & \prod_{e^*} \int dg_{e^*} \prod_e \frac{(1+ \epsilon(G_e))}{2} \prod_{e} \int d \vec{x}_e \prod_{e \in \Gamma} \int d \vec{u}_e  \,\, e^{i \, S} \\
& = &  \int \prod_{e^*} dg_{e^*} \, \prod_e \frac{(1+ \epsilon(G_e))}{2}\, \tilde{Z}_p \tilde{Z}_c
\end{eqnarray*}

Where,
\begin{eqnarray*}
\tilde{Z}_p & = & \prod_{e \notin \Gamma}\int d \vec{x}_e  \,\, e^{i S_p} \\
\tilde{Z}_c &  = & \prod_{e \in \Gamma} \int d \vec{x}_e \, d \vec{u}_e \,\, e^{i S_c}
\end{eqnarray*}

The above separation of the partition function is possible because the action (\ref{eq:traced}) is factorized per edge (dual face) as far as the $\vec{x}_e$ and $\vec{u}_e$ variables are concerned. We want to do the integrals over $\vec{x}_e$'s and the $\vec{u}_e$'s. The treatment of $\tilde{Z}_p$ is exactly the same as it was before, and was shown in the previous section.\\

We move on to $\tilde{Z}_c$. Now $\tilde{Z}_c$ can be written as a product of contributions coming from each edge of $\Gamma$. 

\[
\tilde{Z}_c = \prod_{e \in \Gamma} \tilde{Z}^e_c
\]

where

\[
\tilde{Z}^e_c = \int d \vec{x}_e \, d\vec{u}_e \, e^{i S_c^e}
\]
 
where $S_c^e$ is the part of the action (\ref{eq:traced}) coming from edge e. We shall concentrate on only one edge of $\Gamma$ and, consequently, drop the 'e' labels on the variables from the formulas below. 

\begin{equation}
\label{eq:Z_c^e}
\tilde{Z}_c^e = \int d \vec{x} \, d \vec{u} \,  e^{- i  \big [ Cos(\tilde{m}) Sin(\theta) \vec{x} \cdot \vec{n} + Sin(\tilde{m}) Cos(\theta) \vec{x} \cdot \vec{u} - Sin(\tilde{m}) Sin(\theta) \vec{x} \cdot ( \vec{n} \times \vec{u}) \big ]   }
\end{equation}

We want to do the integrals over $\vec{x}$ and $\vec{u}$ now. What are the ranges of integration? For $\vec{u}$ the answer is simple: it is $S^2\simeq SU(2)/U(1)$, where the $U(1)$ chosen is the subgroup of $SU(2)$ to which $h=e^{m J_0}$ belongs. What should the range of integration for $\vec{x}$ be? From considerations in the previous section, we expect it to be half of $\mathbb{R}^3$, the question is, which half?\\ 

We know that, in the continuum, the component of E along the trajectory of the particle, is equal to the proper time. This means that the component of E along the momentum of the particle gives the proper time, and it is this component that we must restrict to positive values. We have two potential vectors to call 'momentum', F and p (see (\ref{eq:classical coupled})). At the classical level, there is no difference between F and p, since the equations of motion (\ref{eqnarray:curvature}) set them equal to each other. It might seem at first glance that p is the right momentum to use. However, it turns out that, since we are interested in the quantum theory, F is the right choice. This is because the magnitude of p is fixed to be m, i.e. p is already on shell. But we know that in quantum mechanics the momentum of the particle can be anything it wants and is not restricted to the classical value. Thus, we conclude that F is the more suitable candidate for momentum. Of course, this is confirmed by the whole body of knowledge about quantum particles coupled to 3d quantum gravity, as can be seen in \cite{hooft, matschullwelling, desousa, PRI}, as well as by the results we are about to present.\\

It follows from the above that we must restrict the component of E along the F to positive values only. Going through the same arguments as in the previous section, it follows that in the discrete case this means restricting the integration in the partition function in such a way so that $\vec{x} \cdot \vec{n} \geq 0$. Notice that this is {\it exactly} the same prescription that we have used in the pure quantum gravity case. This confirms once more that there is a universal (i.e. valid in presence as well as in absence of matter) restriction on the {\it geometric} variables that produces the correct {\it causal} quantum amplitudes we are after.\\

The upshot of the above is that we do the $\vec{x}$ integral over the half space for which $\vec{x} \cdot \vec{n} \geq 0$, remembering\footnote{See Appendix B.} that this implies that the $\vec{n}$ has to be complexified, or equivalently that the $z = \vec{n} \cdot \vec{u}$ is complex (hence the suggestive notation). The result of integrating (\ref{eq:Z_c^e}) over this range of $\vec{x}$ gives\footnote{See Appendix B again. We have implicitly divided by the constant $C_{\delta}$ from the previous section. Again, the factor of half is needed if one wants to recover the coupled Ponzano-Regge with usual normalization for the amplitudes (no numerical constants in front of the delta function).}
\begin{equation}
\label{eq:distro2}
\tilde{Z}^e_c = \int d \vec{u} \, \frac{1}{4} \frac{i}{(\sqrt{b^2})^3}
\end{equation}

where
\[
b^2 = - (\vec{n} \cdot \vec{u})^2 Sin^2(\theta) Sin^2(\tilde{m}) + 2 (\vec{n} \cdot \vec{u}) \, Sin(\theta) Cos(\theta) Sin(\tilde{m}) Cos(\tilde{m}) + 1- Cos^2(\tilde{m}) Cos^2(\theta) .
\]

Substituting this expression for b into (\ref{eq:distro2}), and using spherical coordinates to do one angular integral we get

\begin{equation}
\label{eq:distro3}
\frac{1}{4} \int_{-1+ i \epsilon}^{1+i \epsilon} dz \, i \, \Big [ \sqrt{-
    Sin^2(\theta) Sin^2(\tilde{m}) z^2 + 2 Sin(\theta) Cos(\theta)
    Sin(\tilde{m}) Cos(\tilde{m}) z + 1- Cos^2(\tilde{m})
    Cos^2(\theta)}\Big ]^{-3} \label{eq:integralz}
\end{equation}

Before we analyze this integral, let us simplify it somewhat. So, the integral above is equal to

\be
\frac{1}{4} \int_{-1+ i \epsilon}^{1+ i \epsilon} \frac{i \, dz}{\bigg (1- Sin^2(\theta) Sin^2(\tilde{m}) \Big [ z - Cot(\theta) Cot(\tilde{m}) \Big]^2 \bigg )^{\frac{3}{2}}} 
=\frac{i}{4 Sin(\theta) Sin(\tilde{m})} \int_{-Cos(\theta-\tilde{m}) + i \epsilon}^{-Cos(\theta +\tilde{m})+ i \epsilon} \frac{dw}{\bigg ( 1- w^2 \bigg )^{\frac{3}{2}}} \label{eqnarray:complex integral}
\ee

Once more, it is crucial to keep in mind that the above is a
\textit{complex} integral, otherwise the step from
(\ref{eq:Z_c^e}) to (\ref{eq:distro2}) is simply incorrect. This
comes from the need to complexify the group element (its
parameters, and thus $\vec{n}$) in order to make sense of the integral over
$\vec{x}$ that gives rise to (\ref{eq:integralz}). In other
words, we are not trying to evaluate a real integral using complex
methods. The integral \textbf{\textit{is}} complex to start with, and
forcing it into the real domain creates numerous problems, which are
simply not there if one deals with it correctly. The main problem with
considering the integral as it were in the real domain is that one
loses analyticity: the real function resulting from performing
(\ref{eq:integralz}) as it was a real integral is not only
discontinuous, which is normal as it has poles, but it does not admit
any analytical continuation even away from the poles.
As such, it cannot be considered as a propagator (Feynman or else) of
any field theory. In fact, as a consequence of the non-analyticity, the action of the theory, in momentum space,
which would have it as its free propagator, would be discontinuous across the point
where the momentum is equal to the mass (classical solution of the
field equations). More precisely, the action switches from the one for
a usual scalar field theory when the momentum is less than the mass,
to an action of a nonlocal theory when the momentum is larger than the
mass. All these problems persist even in the no-gravity ($G\rightarrow
0$) limit.

All these issues are solved automatically once one realizes that
(\ref{eq:distro2}) is a complex integral from the very beginning. Once
this has been done, one has then to specify the contour of
integration, that will depend on the two parameters $\theta$ and $\tilde{m}$.
Moreover, the ambiguity in our case is even greater, since the
integrand is a complex function with branch points, and the natural
domain of integration becomes the associated Riemann surface and not
the complex plane. Thus there are even more possible paths of
integration in this case, since we can go from one sheet to the other.

The upshot of the above is that after taking analyticity into account
there are essentially only two choices of the contours of
integration. One choice gives the usual Feynman propagator as we shall
see. The other seems to give the difference of advanced and retarded
Green functions (instead of the sum, which is the Feynman
propagator), and would correspond to a non-local field theory even in
the abelian limit.
We show the contours of integration on the Riemann surface and on the
complex plane in Appendix C.

If one follows the contours prescribed, one obtains as a result of the z integral,
\[
\frac{Cos(\theta)}{2 Sin(\tilde{m})} \, \frac{i}{Sin(\theta-\tilde{m}) \, Sin(\theta+\tilde{m}) + i \epsilon}
\]

Or, after trigonometric manipulations
\begin{equation}
\label{eq:matter propagator}
\frac{Cos(\theta)}{2 Sin(\tilde{m})} \, \frac{i}{Sin^2(\theta)  -  Sin^2(\tilde{m}) + i \epsilon}
\end{equation}

This last expression, agrees, up to a $Cos(\theta)$ factor which is constant in the abelian limit, with the expression for the nonabelian Feynman propagator put forward in \cite{PRIII}. However, this factor affects the non-abelian, i.e. quantum gravity-dependent, structure of the theory in a very non-trivial way, and will affects the effective non-commutative field theory corresponding to it, as we are going to see in the next section. It is therefore important, and cannot be neglected. Of course, the main reason not to neglect it is that it follows directly and uniquely from the discrete path integral we have defined and computed, so it is a necessary part of the definition of a spin foam model for 3d quantum gravity. Also, let us stress that, had we used or obtained delta functions over the group enforcing the classical equations of motion, then this factor would have been fully equivalent to a constant normalization factor $Cos(m)$, indeed present in the normalization used in \cite{PRIII}, and thus physically irrelevant. This is the effect of the ambiguity in the re-construction of the Feynman propagator from the Hadamard one that we had mentioned in section II. There is also a constant factor equal to $\frac{2}{Sin(\tilde{m})}$ coming from every edge of $\Gamma$ (also present in \cite{PRIII}). 

Putting everything together, we arrive thus at the following expression for the partition function for gravity coupled to matter
\begin{equation}
\label{eq:full partition}
Z = \int \prod_{e^* \in \Delta^*}   d g_{e^*} \prod_e \frac{(1+ \epsilon(G_e))}{2} \prod_{e \notin \Gamma} \frac{1}{4} \frac{i}{(Sin(\theta_e) + i \epsilon_e)^3} \prod_{e \in \Gamma}\frac{Cos(\theta_e)}{2 Sin(\tilde{m}_e)} \,  \frac{i}{Sin^2(\theta_e) -Sin^2(\tilde{m}_e) + i \epsilon_e}
\end{equation}

The properties of the above expression are very similar to those of (\ref{eq:pure partition}). In particular
\begin{itemize}
\item The amplitude is sharply peaked around the classical configuration, which is $\theta_e =0$ for $e \notin \Gamma$, and $\theta_e = \tilde{m}_e$ for $e \in \Gamma$, in agreement with one's expectations.
\item It is clear that (\ref{eq:full partition}) changes into the complex conjugate of itself, if one switches from integrating over the $\vec{x}$'s that make a positive dot product with appropriate vectors to those which a make a negative product with those same vectors. This is precisely what one would expect from continuum considerations.
\item The model is not triangulation independent, essentially because the particle-free part of the amplitude is not\footnote{Since we chose the triangulation to be adapted to the particle graph, to start with, there is not much sense in asking whether the model is triangulation independent \textit{at} the location of the particle (Feynman graph).}.
\end{itemize}

One can of course give a spin foam presentation of the above partition function. This is given by:  
\[
Z = \sum_{labellings} \, \, \prod_{e\notin \Gamma} (2 j_e + 1)\,\prod_{e \in \Gamma} K_{j_e}(\tilde{m}_e , \epsilon) \prod_{tetrahedra \in \Delta} \{ 6j \}
\]

The sum goes over all labellings of the faces of the dual two-complex by representations of SU(2). In the above $\{ 6j \}$ denotes the usual 6-j symbol, and the $K_{j}(\epsilon)$'s are given by:

\begin{eqnarray*}
\lefteqn{\frac{(1+ \epsilon(G))}{2}   \frac{Cos(\theta)}{2 Sin(\tilde{m}_e)} \,  \frac{i}{Sin^2(\theta) -Sin^2(\tilde{m}_e) + i \epsilon_e} \, = \, \sum_j \, K_j (\tilde{m}_e, \epsilon) \chi_j(\theta)} \\
&& K_j(\tilde{m}_e,\epsilon) \, = \, \frac{2}{\pi}  \int_0^{\frac{\pi}{2}} \frac{Cos(\theta)}{2 Sin(\tilde{m}_e)} \,  \frac{i}{Sin^2(\theta) -Sin^2(\tilde{m}_e) + i \epsilon_e} \, \chi_j(\theta)  \, Sin^2(\theta) \, d \theta = \frac{ i\,e^{-i (2 j + 1) (\tilde{m}_e- i \epsilon)}}{2 Sin(\tilde{m}_e)} + i^{(2 j+1)} F_j  \\ && 
F_j \, = \frac{1}{2 \pi Sin(\tilde{m}_e)}  \int_{-1}^{1} \frac{(x^4-1) \, x^{2 j}}{(x^2 + e^{2 i m}) (x^2 + e^{-2 i m}) } dx
\end{eqnarray*}

The $F_j$'s can in fact be evaluated explicitly  in terms of elementary functions (which can be neatly repackaged into the hypergeometric functions), however what is important about them is that:
\begin{itemize}
\item For half-integer representations (j is a half-integer), $F_j$ is zero. This is immediate from the fact that the integrand is odd in this case.
\item $F_j$ is manifestly real.

Because of these two properties, the sum of $K_j$ and its complex conjugate, i.e. the real part of the above coefficients, is simply given exactly by the character  $\chi_j(e^{\tilde{m}_e J_0})$, so that the sum of the whole partition function with its complex conjugate gives back {\it exactly} the usual coupled Ponzano-Regge model, as one expects. This has to be compared with the proposal for the Feynman propagator in \cite{PRIIIbis}, where such correspondence is realized only modulo proportionality factors (functions of $\tilde{m}_e$) that are instead fixed uniquely from first principles in our procedure.   
\end{itemize}
It is very interesting to look at the quantum field theory (abelian) limit of (\ref{eq:full partition}), along the lines of \cite{PRIII}. We turn our attention to this task now.

\section{The QFT limit}

We want to see now how (\ref{eq:full partition}) compares to the evaluation of the usual Feynman graphs in flat spacetime. This will imply first the re-writing of the above partition function, in the appropriate {\it approximation}, to be explained below, as the evaluation of the Feynman diagrams of a non-commutative field theory that takes into account the quantum gravity induced modifications on the propagation of matter {\it only}; second, it will imply checking that the same partition function reduces to the evaluation of the Feynman graphs of {\it ordinary} scalar field theory in flat space in the no-gravity or $G\rightarrow 0$ limit. 

In other words, we want to concentrate on the matter part of (\ref{eq:full partition}) and 'integrate out' the gravity degrees of freedom. To do that note that by far the dominant contribution to the partition function from the no-particle edges, comes from the  configurations which are flat on those edges. This means that, as far as the particle edges are concerned, the only contributions that are relevant are those for which the connection is flat everywhere except at the particle graph. Of course, this is an approximation, we are replacing the fully quantum partition function, with the one in which gravity is flat (classical, on-shell), everywhere where there is no matter.\\

To turn the above into formulas, note that action (\ref{eq:classical coupled}) can be written as
\[
S = \int_M Tr( E \wedge F) - \sum_i \int_{\gamma_i} Tr( E\, p_i) = S_p(g_{\mu \nu}) + S_c(g_{\mu \nu},p)
\]

where $g_{\mu \nu}$ stands of course for the metric. We have switched to the second order formulation to make the physical point more transparent. Now, what we want to do is to approximate the partition function in the following way:

\begin{eqnarray*}
Z & = & \int \mathcal{D} g_{\mu \nu} \,  \prod_i \int \mathcal{D} p_i  \, e^{i S} \,=\,  \int \mathcal{D}g_{\mu \nu} \, \prod_i \int \mathcal{D} p_i \, e^{i S_p (g_{\mu \nu})} \, e^{i \sum_i S_c (g_{\mu \nu}, p_i)} \\
& = & \int \mathcal{D} g_{\mu \nu}^p \, \mathcal{D}g_{\mu \nu}^c \,  \prod_i \int\mathcal{D} p_i \,  e^{i S_p (g_{\mu \nu})} \, e^{i \sum_i S_c (g_{\mu \nu}, p_i)} \, \sim \, \int \mathcal{D}g_{\mu \nu}^c \, \prod_i \int\mathcal{D} p_i \,  e^{i S_p(\eta_{\mu \nu})} \, e^{i \sum_i S_c (g_{\mu \nu}, p_i)}\\
\end{eqnarray*}

in the above $g_{\mu \nu}^p$ and $g_{\mu \nu}^c$ denote the metric that appears in the pure gravity or in the coupled part of the action respectively. $\eta_{\mu \nu}$ is the Minkowski metric.\footnote{It may help to note that the discrete analogues of $S_p(g_{\mu \nu})$ and $S_c(g_{\mu \nu}, p_i)$ were given in (\ref{eq:traced}).} \\

Putting the above differently, gravity has a dual role in this model. It has independent (only quantum) degrees of freedom itself, and it couples to and deforms (as we shall see below) the matter action. What we want is to freeze the (quantum) dynamics of gravity for now and concentrate on the deformation it causes in matter propagation.\\

Keeping the above in mind, we replace the pure-gravity amplitudes in (\ref{eq:full partition}) with delta functions, obtaining\footnote{This expression is exactly the one obtained in \cite{PRIII}, except for the Cosine factor. We follow the same normalization used there, dividing by the $\frac{1}{2 Sin(m_e)}$ factor.}
\begin{equation}
\label{eq:simple partition}
Z =  \int \prod_{e^* \in \Delta^*}   d g_{e^*} \frac{(1+\epsilon(G_e))}{2} \prod_{e \notin \Gamma} \delta(G_e)   \prod_{e \in \Gamma} Cos(\theta_e) \, \frac{i}{Sin^2(\theta_e) -Sin^2(\tilde{m}_e) + i \epsilon_e}
\end{equation}

as we shall see below, (\ref{eq:simple partition}) can be thought of as a Feynman diagram evaluation of a noncommutative field theory. Since we will be interested in taking the limit in which we recover the usual, flat-space QFT, we must put Newton's constant G back into the equations. \\

To see where the G should go recall that, as was mentioned in section II, $\tilde{m}_e$ is the deficit angle associated to the mass $m_e$. More precisely, comparison with the classical solution \cite{hooft} reveals that

\[
\textrm{Deficit Angle} = \tilde{m}_e = 4 \pi G m_e = \kappa m_e
\]

Thus we have to replace everywhere $\tilde{m}_e$ with $\kappa m_e$. It is natural then to replace $\theta_e$ everywhere with $\kappa \theta_e$, and interpret the equation above as an equation between momenta\footnote{This, in fact, is a highly nontrivial step which was first done in \cite{PRIII}, and although we are trying to make it look plausible, its real justification is an entirely \textit{a posteriori} one: \textit{if} one does this, then the usual quantum field theory is obtained in the $\kappa \arr 0$ limit.}
\[
\kappa \, \, \textrm{Momentum flowing on edge e} = \kappa \, \, \textrm{Mass on edge e}
\]

This reinterpretation is clearest on-shell, but we will retain this interpretation off-shell as well, thus interpreting the gravitational holonomy corresponding to $\theta$ as the true quantum (and thus off-shell) momentum of the particle. As we have stressed above, this is of course consistent with what we know \cite{hooft, desousa,matschullwelling,PRI} about quantum point particles coupled to 3d quantum gravity, as well as with our own results. So, we shall parametrize our group elements as
\begin{equation}
\label{eq:first param}
g = Cos(\kappa \theta) 1 + \vec{J} \cdot \vec{n} Sin(\kappa \theta)
\end{equation}

The above parametrization is a very physically transparent one, and it is most useful for taking the $\kappa \arr 0$ limit. To get the noncommutative field theory, we shall use however a slightly different way of representing the group elements given by 
\[
g = \sqrt{1 - \kappa^2 | \vec{P} |^2}1 + \kappa \vec{P} \cdot \vec{J}.
\]
In doing so we have introduced the non-commutative momentum coordinates $\vec{P}$, valued in $\mathcal{B}_\kappa^3$, the 3-ball with radius $1/\kappa$ in $\mathbb{R}^3$ \footnote{Following the definition, the integration measures for $g$ and $P$ are related by: $$ \int_{SU(2)} dg = \int_{\mathcal{B}_{\kappa}^3}\frac{\kappa^3\,d^3 P}{\sqrt{1-\kappa^2\mid \vec{P}\mid^2}}$$}, previously mentioned in section II and first used in \cite{PRIII}.

Now, using the star product introduced in \cite{PRIII}, given by
\[
e^{\frac{i}{2 \kappa} Tr(X g_1)} \star e^{\frac{i}{2 \kappa} Tr (X g_2)} = e^{\frac{i}{2 \kappa} Tr(X g_1 g_2)}
\]

and the 'Fourier' transform naturally associated to the above product\footnote{X is an $\mathfrak{su}\,$(2) element, while $g_1, g_2$ are SU(2) group elements. Also, the minus in the exponent is purely a matter of convention. The choice here is made so that the formulas agree with those in \cite{PRIII}.}
\[
f(X) = \int d g \, e^{-\frac{i}{2 \kappa} Tr(X g)} \tilde{f}(g)
\]

it can be shown, using the very same procedure used in \cite{PRIII}, that (\ref{eq:simple partition}) can be rewritten as\footnote{$|E|$ as before is the number of the edges of the triangulation.}
\begin{equation}
\label{eq:noncom partition}
\frac{1}{\kappa^{2|E|}} \int \prod_v \frac{d^3 \vec{X}_v}{8 \pi \kappa^3} \int \prod_e \frac{\kappa^3 d^3 \vec{P}_e}{\pi^2} \frac{i}{|\vec{P}_e|^2 - \frac{sin^2(m_e \kappa)}{\kappa^2} + i \epsilon_e} \prod_v \Big ( 
\bigstar_{e \in \partial v} e^{i \vec{X_v}\cdot \vec{P}_e}  \Big )
\end{equation}

The integrals over $\vec{X}_v$'s give delta functions, enforcing momentum conservation at the vertices. The expression that one gets is

\[
\frac{1}{\kappa^{2|E|}}  \prod_e \int \frac{\kappa^3 d^3 \vec{P}_e}{\pi^2} \prod_{e\in\Gamma}\frac{i}{|\vec{P}_e|^2 - \frac{sin^2(m_e \kappa)}{\kappa^2} + i \epsilon_e} \prod_{v\in\Gamma}\delta{\big(\oplus_{e \in \partial v} \vec{P}_e \big)}
\]

It is clear that the above expression has {\bf exactly} the form of a usual Feynman diagram evaluation, but in terms of the non-commutative momenta $P$, i.e. for a non-commutative field theory that we are going to write explicitly in the following. It looks the same as the one obtained in \cite{PRIII, PRIIIbis}, except for the integration measure over the $P$s, that is given here by an apparently usual $\kappa^3 d^3P$, while was given by $dg=\frac{\kappa^3 d^3P}{\sqrt{1-\kappa^2\mid \vec{P}\mid^2}}$ in \cite{PRIII,PRIIIbis}. It is also important to notice that the fact that the (noncommutative) momentum is conserved follows here from moving from (\ref{eq:full partition}) to (\ref{eq:simple partition}). In other words, conservation of momentum is only valid when gravity degrees of freedom are 'frozen', and the dynamics of pure quantum gravity as encoded in the partition function (\ref{eq:full partition}) is neglected and the partition function itself is evaluated at the dominant contribution (flat spacetime) only. This means that momentum conservation, even at the non-commutative level, is only {\it approximately} satisfied in the causal theory.

 Inverting the expression for the Feynman propagator used in (\ref{eq:noncom partition}), one can see \cite{PRIII} that the action that generates Feynman diagrams of the form\footnote{If one restricts to trivalent graphs.} (\ref{eq:noncom partition}) is given by the following scalar noncommutative field theory action

\[
S(\phi) = \frac{1}{2} \int dg\frac{ \Big (   P^2(g) - \frac{Sin^2(\kappa m)}{\kappa^2}    \Big )}{\sqrt{1 - \kappa^2 | \vec{P} |^2}} \tilde{\phi}(g) \tilde{\phi}(g^{-1}) + \frac{\lambda}{3!} \int dg_1 dg_2 dg_3 \delta(g_1 g_2 g_3) \tilde{\phi}(g_1) \tilde{\phi}(g_2) \tilde{\phi}(g_3)
\]

The coupling constant $\lambda$ has been introduced to make the action above look similar to the familiar $\phi^3$ theory. 

The position space version of this action is given by: 

\begin{equation}
\label{eq:position action}
S(\phi) = \frac{1}{8 \pi \kappa^3} \int d^3 x \Bigg [     \frac{1}{2} \Big( \phi \star \Big[\frac{\nabla - \frac{Sin^2(\kappa m)}{\kappa^2}}{\sqrt{1 - \kappa^2 \partial^2}}        \Big]  \phi (x)     \Big)   + \frac{\lambda}{3 !} \Big ( 
 \phi \star \phi \star \phi  \Big ) (x)   \Bigg ]
\end{equation}
where$\nabla$ is the Laplacian on $\mathbb{R}^3$. Once more, the action differs, in both momentum and position space, from the one obtained in \cite{PRIII,PRIIIbis}, due to the additional $Cos(\theta)$ factor per edge that is present in our partition function, and necessarily so because it follows uniquely from the discretization of the BF partition function {\it plus} the causality restriction.

\medskip

We shall take now the $\kappa \simeq G \arr 0$ limit of (\ref{eq:simple partition}). Using the first parametrization (\ref{eq:first param}), it is obvious that integration over the group SU(2) is replaced by integration over its Lie algebra $\mathfrak{su}\,(2)\simeq \mathbb{R}^3$. 
\[
\int_{SU(2)} dg_{e^*}  \arr  \kappa^3 \int_{\mathbb{R}^3} d^3 \vec{p_{e^*}}
\]
Where we use the notation $\vec{p} = \theta \vec{n}$. It is clear that with this definition, the group elements (\ref{eqnarray:group element}) can be expanded in powers of $\kappa$ as
\[
G_e = \big ( 1  -  \kappa^2 \frac{|\vec{p_e}|^2}{2} \big ) + \kappa \vec{p_e}\cdot \vec{J} + \ldots
\]

Keeping the above in mind, and using the same techniques as in \cite{PRIII}, in particular using the triangulation independence of the approximate particle-free part of the partition function, ensured by the use of delta functions over the holonomies of the connection, we see that the abelian limit of the full partition function (\ref{eq:full partition}) is given  by:
\begin{equation}
\label{eq:abelian limit}
\lim_{\kappa \arr 0} Z =  \prod_{e \in \Gamma} \int d \vec{p_{e}} \prod_{v\in \Gamma}\, \delta \left(\sum_{e\mid v}\vec{p_e}\right) \prod_{e \in \Gamma}\, \frac{i}{ |\vec{p_e}|^2 - m^2 + i \epsilon_e}
\end{equation}

The equation above is almost exactly the same as the expression obtained in \cite{PRIII} with the sole (welcome) difference in that the amplitudes on the particle edges are the Feynman propagators instead of the Hadamard ones, i.e. (\ref{eq:abelian limit}) is equal to the evaluation of the usual Feynman diagram associated to the graph $\Gamma$ in flat spacetime for a scalar field.\footnote{At the level of the action, it is very easy to see that (\ref{eq:position action}) reduces to the usual scalar field theory action in the limit $\kappa \arr 0$, since the star product goes to pointwise multiplication in this limit, and the derivative factor in the denominator of the kinetic term drops out.} We see that, as anticipated, the abelian limit of the causal coupled partition function is insensitive to the edge factor $Cos(\theta)$ that marked the difference between the non-commutative Feynman propagator as uniquely obtained from the causal restriction of the BF partition function and the one used in \cite{PRIII}.

Let us also emphasize once more that the noncommutative analogue of (\ref{eq:abelian limit}), is \textit{not} exactly the effective field theory coming from the full partition function. What it \textit{does} correspond to is the theory in which instead of the $\frac{i}{(Sin(\theta)+i \epsilon)^3}$ factors for the particle-free edges, one has delta functions. In other words, it is the theory in which gravity is on-shell away from the particles. It is clear that the off-shell partition function (\ref{eq:full partition}) is not exactly equal to the noncommutative field theory written above, because of the 'virtual' gravity degrees of freedom present\footnote{This makes equation (\ref{eq:full partition}) dependent on the triangulation chosen, and thus on extra information not contained in the Feynman graph $\Gamma$ alone. This last point makes unlikely, in our opinion, although not ruled out that there exists  a different non-commutative field theory that gives the full true partition function as the evaluation of its Feynman graphs.}.

\section{Conclusions and outlook}

Let us summarize our results. We have shown what is the correct discrete counterpart of the causality condition that reduces BF theory to 3d quantum gravity in the Lagrangian setting, i.e. the restriction on the geometric variables that imposes on the one hand the correct dependence of the path integral on the spacetime orientation, and, on the other hand, realizes the correct 'off-shell' (with respect to the canonical Hamiltonian constraint) propagation of gravity degrees of freedom. We have constructed the corresponding {\it causal} spin foam model for pure 3d quantum gravity, and discussed its properties. We have then coupled matter to the model, along the lines of \cite{PRI}, and derived, through a rigorous discretization and evaluation procedure, a {\it causal} coupled spin foam model for 3d gravity and scalar matter (the model can however be trivially extended to include spin degrees of freedom, whose coupling is not affected by the causality restriction). Then, we have shown how the causal model admits a physically motivated {\it approximation} corresponding to approximating the integral over pure gravity degrees of freedom with its dominant contribution, represented by the flat spacetime geometry. This approximated model has then been re-written as the Feynman evaluation of a Feynman graph for a non-commutative scalar quantum field theory, encoding fully the matter degrees of freedom plus the quantum gravity corrections to matter propagation, while the un-approximated model encodes also the quantum gravity corrections to flat space momentum conservation. The amplitudes so obtained involve non-commutative Feynman propagators, as one would expect. No ambiguity at all arises in re-writing the approximate partition function in terms of such Feynman evaluations, and everything follows uniquely from the original discretization of the causally restricted partition function. Finally, we have extracted, from the form of the non-commutative Feynman amplitudes, the unique non-commutative field theory action producing them in perturbative expansion, both in momentum and configuration space., and shown that both the action and the corresponding Feynman amplitudes reduce to the correct usual Feynman amplitudes for a scalar field theory in flat commutative space in the no-gravity limit.  

\medskip
Let us now conclude by mentioning only some of the many possible uses and further developments of the results we have presented. 

We can summarize what we have achieved as the realization, in a discrete context and in a spin foam language, of the path integral or sum-over-histories idea for 3d quantum gravity coupled to matter fields, in the Lagrangian framework. One can then use this path integral for computing quantum gravity transition amplitudes and observables, being assured that the formalism correctly encodes causality and orientation dependence restrictions. However, it is maybe of more direct physical interest to compute the effects of quantum gravity corrections to matter scattering amplitudes, now that the unique and correct expression for the quantum gravity modified Feynman amplitudes for matter fields have been derived.

There are several reasons, however, for believing that only a group field theory context would render the use of the above amplitudes truly physically meaningful (as far as 3d quantum gravity can be physically meaningful, of course). One obvious one is the triangulation dependence of the causal model, both for pure gravity and matter coupling. This triangulation dependence has to be expected, as we have discussed, since 3d gravity lacks local degrees of freedom only on-shell, i.e. only after the Hamiltonian constraint has been imposed, that is what causality prevents us from doing, in a sense. On the other hand, this triangulation dependence is still physically troubling, as it indicates that we cannot restrict ourselves to a fixed choice of simplicial complex on which to define our theory, as this implies necessarily a truncation of the degrees of freedom that are captured by our model. The natural way to overcome this limitation, that is also the only known way in the physical 4d case, is to pass to a group field theory formulation, and to generate our spin foam model within a perturbative expansion of a group field theory that has the interpretation of a quantum field theory of simplicial geometry \cite{ioGFT,laurentGFT,GFTbook}. On the one hand this would extend greatly the scope and context of our model, given that GFTs include in their perturbative expansion a sum over different topologies as well as different triangulations and that at the same time point towards possible non-perturbative results as well \cite{GFTbook}. On the other hand, going to a GFT context would first of all solve the issue with triangulation dependence, as we said, but also provide our causal model with an even more natural role. As we have discussed briefly in the introduction, in fact, the need for an \lq off-shell\rq propagation of quantum gravity degrees of freedom and thus for the un-symmetrized Lagrangian sum-over-histories, as opposed to the symmetrized one realizing just the projection onto physical states, becomes absolutely apparent in a 3rd quantized context, i.e. that of a true field theory formulation of quantum gravity. This is exactly what is provided by the GFT framework for simplicial quantum gravity and spin foam models.\\

As mentioned, a proposal for a class of generalized group field theories, corresponding to {\it causal} spin foam models, in any spacetime dimension and signature, exists \cite{generalised}, and can be specialized to the 3d case in a rather straightforward manner. The first thing to check, therefore, is whether our causal model can be seen as arising from this class of GFTs. If it does not, the modification of the generalized GFT formalism of \cite{generalised} that instead does the job should be sought for. Having at hand such a GFT model, the coupling of matter fields for it, along the lines of \cite{iojimmy}, has to be investigated. Only after this, we think, we can truly start tackling physical issues on the basis of the causal model we have constructed.\\   

Of course, true physics needs a 4-dimensional setting, and thus the main aim should be to work in the context of 4-dimensional quantum gravity. However, as it is well known, no consensus has been reached yet concerning the correctness of any specific 4-dimensional spin foam model, and the analysis of the proposed ones has still to progress past the initial stage. This is true also at the group field theory level. Concerning the issues that have interested us in the present work, i.e. the issue of causality in quantum gravity, and in gravity coupled to matter fields, this is not really dependent on the spacetime dimension. Therefore we expect our procedure and results to be of direct application and extension to any 4-dimensional spin foam model. In particular, if the 3d case can be handled successfully within the generalized GFT context of \cite{generalised}, this would imply almost automatically that the same formalism is capable of implementing correctly causality restrictions in any dimension. Much more work is also needed to understand the coupling of matter fields in the 4d context; however, once again, our procedure and result would most likely be of direct application to any spin foam model of 4d quantum gravity coupled to matter that makes an explicit connection with classical continuum and discrete actions. An interesting testing ground would be, for example, the model for topological gravity coupled to string- and brane-like excitations investigated in \cite{branes}. 

\section*{Acknowledgements}
We would like to thank J. Ambjorn, S. Hartnoll, R. Loll, J. Ryan, and especially L. Freidel and E. Livine for comments and suggestions.

\section*{Appendix A}

Here we collect a few facts about the group SU(2) and its Lie algebra $\mathfrak{su}\,$(2) that we used in the text.\\
The basis for the Lie algebra is $J_i = i \sigma_i$, where $\sigma_i$ are the Pauli matrices. Thus
\begin{displaymath}
J_0  = \left( \begin{array}{ccc} 
i & 0 \\
0  & -i 
\end{array} \right)    
\, \, , \, \, J_1 = \left( \begin{array}{ccc} 
0 & i \\
i  & 0 
\end{array} \right)   
\, \, , \, \, J_2 = \left( \begin{array}{ccc} 
0 & 1 \\
-1  & 0 
\end{array} \right)   
\end{displaymath}

These obey
\begin{eqnarray*}
[ J_i \, , \, J_j]    & = & - 2 \epsilon_{ijk} \, J_k   \\
& & \\
\{ J_i \, , \, J_j \}  & = & - 2 \delta_{ij} \\
& & \\
Tr( J_i) & = & 0
\end{eqnarray*}

From which it follows that
\[
\vec{J} \cdot \vec{a} \, \vec{J} \cdot \vec{b}  =  - \vec{a} \cdot \vec{b}  -  \vec{J} \cdot ( \vec{a} \times \vec{b})
\]

The last equation has been used in converting the actions we dealt with ((\ref{eq:discrete action}), (\ref{eq:coupled discretized})) to vector form.\\

Using the exponential map, the general SU(2) group element can be written as
\[
e^{\theta \vec{n} \cdot \vec{J}} =   Cos(\theta) 1 + \vec{J} \cdot \vec{n} Sin(\theta)
\]

Where $\vec{n}$ is a unit vector, and $\theta$ is a real parameter which goes from 0 to $\pi$.

\section*{Appendix B}

Here we prove somewhat more rigorously equations (\ref{eqnarray:distro2}) and (\ref{eq:distro2}). The proof and notation below are from \cite{distributions}.
To begin with we define:
\begin{itemize}
\item A cone in $\mathbb{R}^3$, is a set C such that if $x \in C$ then $\lambda \, x \in C$, $\forall \lambda \geq 0$. We shall restrict our attention to acute cones. The simplest definition for an acute cone is that which states that a cone is acute if its intersection with its inverse through the origin is empty.
\item The cone conjugate to C, is the cone $C^*$ such that $\xi \in C^*$ if $\xi \cdot y \geq 0 \, , \forall y \in C$.
\item The tubular region $T^C$ associated to the cone C is the subset of $\mathbb{C}^3$, which is given by
\[
T^C = \mathbb{R}^3 + i C = [ z = x + i y :  x \in \mathbb{R}^3 \, , \, y \in C]
\]
\end{itemize}

We let C now be a very 'narrow' cone along the z-axis. In other words in polar coordinates
\[
C = (r, \theta, \phi) \in \mathbb{R}^3 \, \, \textrm{with}\, \, 0 \leq \theta < \delta \, , \, r > 0
\]
where $\delta$ is a small number (which we will want to take to zero in the end, since what we want is the degenerate case in which C is just the z-axis). Note that $C^*$ is almost the entire half-space with $z \geq 0$ in this case. We claim that
\[
\int_{C^*} e^{i (z \cdot \xi )} d \xi    = 2 i^3 \int_{C^* \cap S^2} \frac{d^2 n }{(z \cdot n)^3}
\]
where $z \in T^C$. The integral over n\footnote{This n is different from the one used in parametrizing the group elements in the main text.} can now be done giving (\ref{eqnarray:distro2}) and (\ref{eq:distro2})

The proof is very simple. It can be shown \cite{distributions} that the left hand side of the equation above is a holomorphic function of z, and it is clear that the right hand side is holomorphic, because the denominator never vanishes\footnote{This is why we needed to 'open' the z-axis slightly.}. As is well-known, if two holomorphic functions agree on an interval, then they are equal everywhere. Thus we will be done if we can show that there's an interval in $T^C$ along which the two sides of the equation are equal. we choose the subset of $T^C$ given by the purely imaginary elements, i.e. we let $z = i y \, , \, x=0$. Then

\begin{eqnarray*}
\int_{C^*} e^{i (z \cdot \xi )} d \xi & = & \int_{C^*} e^{- (y \cdot \xi )} d \xi \\
& & \\
& = &  \int_{C^* \cap S^2} \int_0^\infty e^{-r (y \cdot n)} r^2 dr d^2 n \\
& & \\
& = & \int_{C^* \cap S^2} \frac{d^2 n}{(y \cdot n)^3} \int_0^\infty e^{-u} u^2 du\\
& & \\
& = & 2 i^3 \int_{C^* \cap S^2} \frac{d^2 n }{(z \cdot n)^3} \hspace{2cm} q.e.d.
\end{eqnarray*}

Doing now the integral over n, one gets
\[
\int_{C^*} e^{i (z \cdot \xi )} d \xi = \Big [ - 2 \pi \, tan^2(\frac{\pi}{2} - \delta) \Big ] \frac{i}{z^3}
\]

which coincides with (\ref{eqnarray:distro2}) and (\ref{eq:distro2}).\\

The left hand side of the equation above is called the Cauchy kernel for $T^C$. It is important to realize that for us, only the values for z whose real part is along C are the ones that are relevant. This is because our prescription of integration is dependent on z.

\section*{Appendix C}

Here we show the Riemann surface for the integrand in (\ref{eqnarray:complex integral}), and the contours leading to the Feynman propagator.\\

The Riemann surface for the function that we have, is a two sheeted one (this comes from the square root, of course). There are two branch points, at 1 and at -1. To obtain the surface, imagine that one has two copies of the complex plane with a cut along the real axis joining the two branch points. In fact, to make things easier to visualize, imagine that we add the points at infinity, and consider two Riemann spheres with cuts. Then we should glue them in the usual criss-cross way typical of the square root Riemann surface. It is easy to see that the surface obtained is topologically a sphere\footnote{In fact there are punctures at the locations of the branch points, but these can be filled unambiguosly.}.\\

The Riemann surface for the $\frac{1}{(1-w^2)^{\frac{3}{2}}}$ that we have is shown in the first figure.
\begin{figure}
\centering
\includegraphics[width=0.7\textwidth]{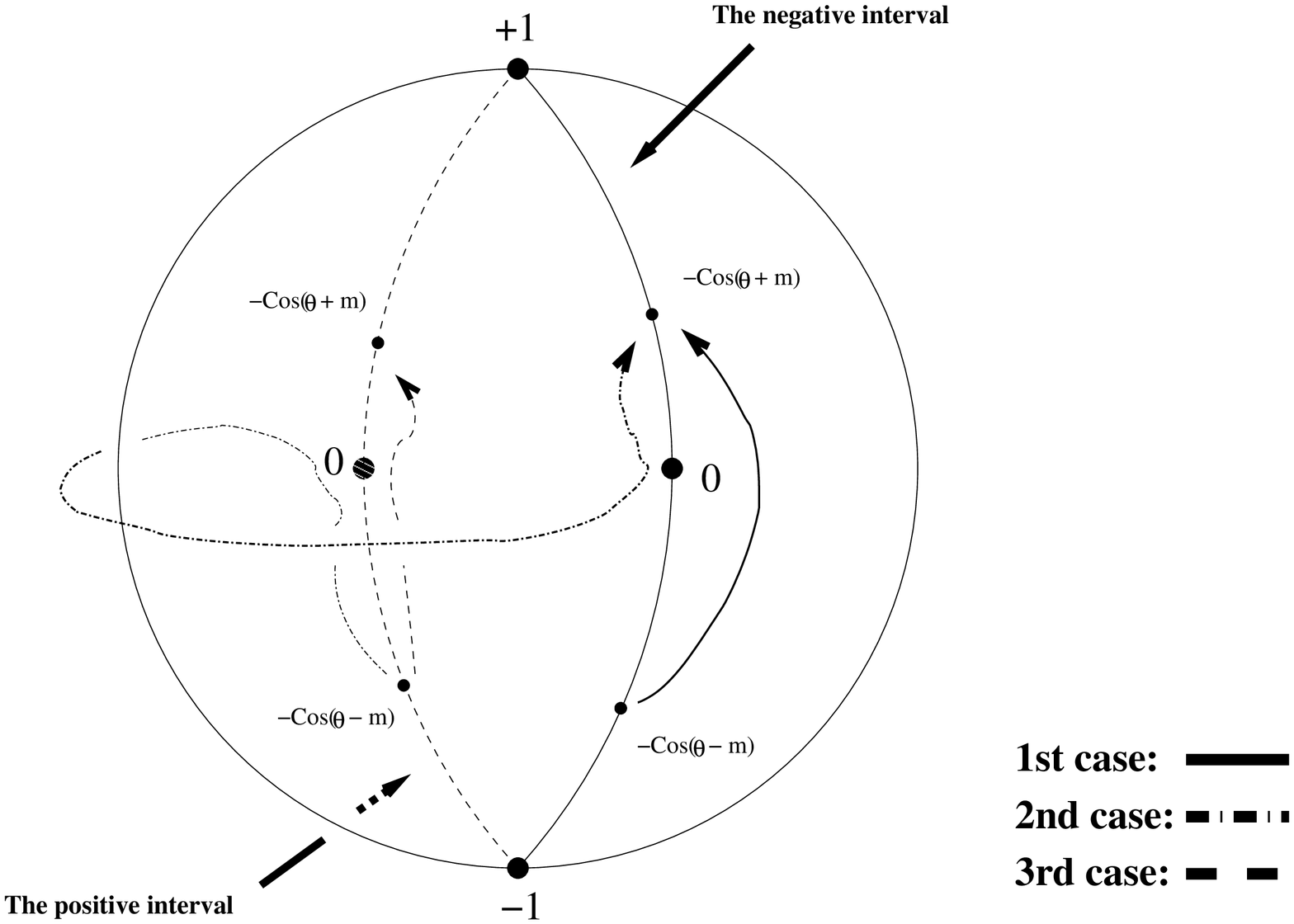}
\caption{\textit{This is the Riemann surface associated with $\frac{1}{(1-w^2)^{\frac{3}{2}}}$. The arrowed lines are the contours of integration in the three cases mentioned in the text.}}
\end{figure}
The meridian that is closer to the viewer, corresponds to the copy of [-1,1] which has negative square roots, while the one that is further away (behind the sphere) is the copy that has positive square roots.\\

There are three distinct cases for our integral:
\begin{itemize}
\item[\textbf{A}-] $\theta \leq m$.
\item[\textbf{B}-] $m \leq \theta \leq \pi -m$.
\item[\textbf{C}-] $\pi -m \leq \theta$.
\end{itemize}

The contours are as shown in the figure\footnote{Needless to say, all contours which can be deformed into each other without passing through singularities or branch points are equivalent.}:
\begin{itemize}
\item[\textbf{A}-] In the first case the contour just follows the 'negative' interval\footnote{One could reverse the conventions, by letting the first case be along the positive interval, and vice versa for the other two cases. This however introduces an overall minus sign into the final expression. We chose this prescription to make the abelian limit agree with usual convention for Feynman propagators.}, i.e. both endpoints are chosen to lie on the copy of [-1,1] which has \textit{negative} square roots.
\item[\textbf{B}-] In this case the contour follows the 'positive' interval from $-Cos(\theta-m)$ to zero, then moves to the 'other' zero, then continues along the 'negative' interval to $-Cos(\theta+m)$.
\item[\textbf{C}-] In the last case, the contour follows the 'positive' interval.
\end{itemize}

A somewhat more conventional picture of the contours is shown in the second figure. Here, the thick line joining the branch points is the cut, and as one approaches it from above one gets positive square roots, while approaching it from below gives the negative ones.

\begin{figure}
\centering
\includegraphics[width=0.7\textwidth]{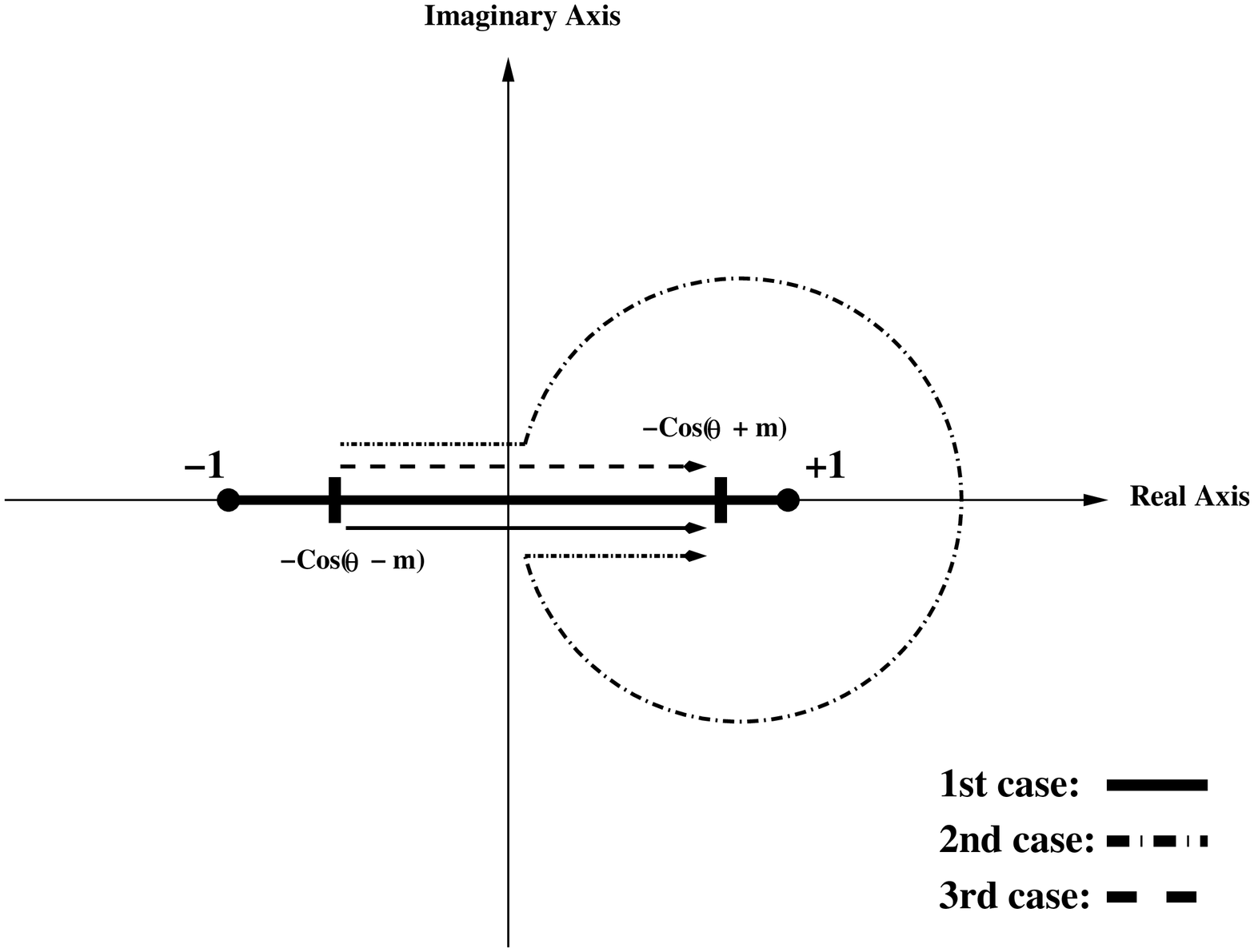}
\caption{\textit{Another picture of the contours of integration. As one approaches the cut (thick line joining -1 and 1) from above one gets positive roots, while approaching from below gives negative ones.}}
\end{figure}

\end{document}